\documentclass[aps,prb,reprint,superscriptaddress,twocolumn,10pt,longbibliography]{revtex4-1}

\usepackage{graphicx}   % Handle images
\usepackage{wrapfig}    % Wrap text around images
\usepackage{float}      % Force image location using "H"
\usepackage{bm}
\usepackage[usenames,dvipsnames,svgnames,table]{xcolor}
\usepackage{epsfig}
\usepackage{amsmath}

\usepackage{amssymb}
\usepackage{array}
\usepackage{nccmath}
\usepackage{dsfont}
\usepackage{array}
\usepackage{cancel}
\usepackage{comment}
\usepackage{hyperref}
\usepackage{cleveref}
\crefname{figure}{Fig.}{Figs.}
\hypersetup{colorlinks=true,breaklinks,linkcolor=blue,urlcolor=blue,citecolor=blue}

\allowdisplaybreaks

\newcommand{\gsexpval}[1]{\ensuremath{\left< #1\right>}}
\def\del{\partial}
\def\tr{\mathop{\hbox{tr}}}

\newcommand{\fb}[1]{\textcolor{red}{#1}}

\begin{document}
\title{Pseudo-Majorana functional renormalization for frustrated XXZ spin-1/2 models with field
or magnetization along the spin-Z direction at finite temperature}
\author{Frederic Bippus}
\affiliation{%
    Institute of Solid State Physics, TU Wien, 1040 Vienna, Austria}%
\affiliation{%
    Munich Center for Quantum Science and Technology (MCQST),  Schellingstr. 4, D-80799 Munich, Germany}%
\affiliation{%
    Department of Physics and Arnold Sommerfeld Center for Theoretical Physics, Ludwig-Maximilians-Universität München, Theresienstr. 37, 80333 Munich, Germany}%

\author{Benedikt Schneider}
\affiliation{%
    Munich Center for Quantum Science and Technology (MCQST),  Schellingstr. 4, D-80799 Munich, Germany}%
\affiliation{%
    Department of Physics and Arnold Sommerfeld Center for Theoretical Physics, Ludwig-Maximilians-Universität München, Theresienstr. 37, 80333 Munich, Germany}%

\author{Bj\"orn Sbierski}
\affiliation{%
    Munich Center for Quantum Science and Technology (MCQST),  Schellingstr. 4, D-80799 Munich, Germany}%
\affiliation{%
    Department of Physics and Arnold Sommerfeld Center for Theoretical Physics, Ludwig-Maximilians-Universität München, Theresienstr. 37, 80333 Munich, Germany}%
\affiliation{Institut für Theoretische Physik, Universit\"at T\"ubingen, Auf der Morgenstelle 14, 72076 T\"ubingen, Germany}
\date{\today}
    
\begin{abstract}
The numerical study of high-dimensional frustrated quantum magnets remains a challenging problem. Here we present an extension of the pseudo-Majorana functional renormalization group to spin-1/2 XXZ type Hamiltonians with field or magnetization along spin-Z direction at finite temperature. We consider a $U(1)$ symmetry-adapted fermionic spin representation and derive the diagrammatic framework and its renormalization group flow equations. We discuss benchmark results and application to two anti-ferromagnetic triangular lattice materials recently studied in experiments with applied magnetic fields: First, we numerically reproduce the magnetization data measured for CeMgAl$_{11}$O$_{19}$ confirming model parameters previously estimated from inelastic neutron spectrum in high fields. Second, we showcase the accuracy of our method by studying the thermal phase transition into the spin solid up-up-down phase of Na$_2$BaCo(PO$_4$)$_2$ in good agreement with experiment.
\end{abstract}
\maketitle

\section{Introduction}

The study of frustrated quantum magnets remains at the forefront of contemporary condensed matter physics, in particular due to their ability to host collective quantum phases of matter like spin liquids \cite{Balents2010,broholmQuantumSpin2020}. 
On the theory side, relevant materials can often be described by relatively compact models like the Heisenberg spin Hamiltonian \cite{auerbachInteractingElectrons1994}. Based on such models, the objective is to compute order parameters, phase diagrams and correlation functions. In the context of solid state quantum magnets, these calculations are usually conducted in thermal equilibrium. However, for frustrated systems like the Heisenberg anti-ferromagnet (AFM) on the triangular lattice, the well established quantum Monte Carlo (QMC) approaches are ineffective due to the sign problem \cite{Sandvik2010}. As an alternative, tensor network methods \cite{Schollwock2011} have been applied both in the ground state (at temperature $T=0$) and, more recently, at finite temperature \cite{ChenExponentialThermalTensor2018,wangSpectroscopyComplextime2024}. Here the bottleneck are usually finite-size effects caused by the non-local entanglement structure.

An alternative approach to frustrated quantum spins are diagrammatic methods which are oblivious to the sign problem or dimensionality. Although originally developed for interacting fermionic (or bosonic) particles \cite{abrikosovMethodsQuantum2012}, variants of the diagrammatic method working directly with spin operators also have a long history\cite{izyumovStatisticalMechanics1988} and currently enjoy renewed interest\cite{kriegExactRenormalization2019,ruckriegelPhaseDiagram2024,schneider2024dipolarordering}. One of the currently most popular diagrammatic methods\cite{Reuther2010} bridges the gap between spin- and fermionic operators using a pseudo-fermionic representation of the spin $S=1/2$ algebra. The resulting interacting fermionic Hamiltonian is then treated with the functional renormalization group (fRG) approach\cite{Kopitz2010,metznerFunctionalRenormalization2012}. For a recent review on this pseudo-fermion fRG (pf-fRG), see Ref.~\onlinecite{Mueller2023}.

While the original pf-fRG was only applied to $T=0$ in the hope to avoid unphysical sectors in the pseudo-fermionic Hilbert space\cite{schneider2024tamingspins}, this restriction was alleviated in 2021 when the fRG was combined with a faithful spin-representation building on pseudo-Majorana operators \cite{Niggemann2021}. The resulting finite-temperature pseudo-Majorana fRG (pm-fRG) has since then been successfully applied for frustrated Heisenberg models with short and long-range interactions \cite{Niggemann2022,Niggemann2023,gonzalez2024,Schneider2024_Tflow,schaden2024xxz} and was also adapted to models with XXZ-type anisotropy \cite{sbierski2023}. 

In this work we generalize the pm-fRG for spin $S=1/2$ XXZ models where time-reversal symmetry is broken by a field in Z-direction, but the $U(1)$ spin-rotation symmetry around this direction remains intact, see Eq.~\eqref{eq_XXZ--Z} below. We dub our method $U(1)$-pm-fRG. Importantly, our framework also applies to the field-free case of a Heisenberg model with $SU(2)$ symmetry spontaneously broken to $U(1)$ by magnetization in spin-Z direction. Similar advances have been put forward\cite{Noculak2024} in the context of pf-fRG at $T=0$.

On the technical level, in Sec.~\ref{Intro_pm-fRG}, we review a $U(1)$-symmetry adapted mixed representation that contains one Majorana- and one complex fermion per lattice site. This representation is also known as ``drone-fermion representation" \cite{Cottam1970,mattis1965,spencer1968}. We discuss the gauge symmetries of the representation and those related to the XXZ--Z Hamiltonian. The associated action and Green's functions are considered in Sec.~\ref{Greens_function_and_action} and we provide the expressions for observables like magnetization and susceptibility in Sec.~\ref{sec_observables}. We derive the (one-loop) flow-equations for $U(1)$-pm-fRG in Sec.~\ref{flow_equations_m_rep}. Various benchmark tests involving small or unfrustrated systems are given in Sec.~\ref{Benchmarking}.

As a showcase application to frustrated systems, in Sec.~\ref{sec:triangular} we apply the $U(1)$-pm-fRG to spin-$1/2$ XXZ--Z models on the triangular lattice inspired by recent experimental studies.
First, we revisit the material CeMgAl$_{11}$O$_{19}$ for which Ref.~\onlinecite{gao2024}  claimed that the coupling parameters are close to a solvable quantum phase transition point in parameter space. While this argument was made on the basis of neutron scattering data in high magnetic fields, we bolster this claim by analyzing low field experimental magnetization data at two different temperatures.

Second, we consider the material Na$_2$BaCo(PO$_4$)$_2$ in an out-of-plane magnetic field. This material was considered recently\cite{Zhong2019,Li2020,gao2024b,Gao2022,Sheng2024} in the context of spin supersolidity\cite{Mila2024,zhu2024,Xiang2024,chen2024}. We determine the critical temperature for the transition in the three-sublattice spin solid up-up-down phase confirming experimental results obtained in Ref.~\onlinecite{Xiang2024}.

We conclude in Sec.~\ref{summary} with summary and outlook. Technical details are relegated to various Appendices.

%%%%%%%%%%%%%%%%%%%%%%%%%%%%%%%%%%%%%%%%%%%%%%%%%%%%%%%%%%%%%%%%%%%%%%%%%%%%%%%%%%%%%
%%%%%%%%%%%%%%%%%%%%%%%%%%%%%%%%%%%%%%%%%%%%%%%%%%%%%%%%%%%%%%%%%%%%%%%%%%%%%%%%%%%%%
\section{XXZ--Z model, Pseudo-Majoranas and the mixed representation}
\label{Intro_pm-fRG}

We consider the XXZ spin $S=1/2$ model with an (optional) external field pointing along the spin-Z direction. The Hamiltonian of this XXZ--Z model reads
\begin{equation}
    H=H_{XXZ}+H_Z \label{eq_XXZ--Z},
\end{equation}
where the first term with spin-spin interactions reads
\begin{equation}\label{eq_XXZ_plain}
	H_{XXZ} = \sum_{i<j}\left[\frac{J_{ij}^{\perp}}{2} \left(S_i^+S_j^- + S_i^-S_j^+\right) + J_{ij}^z S_i^zS_j^z \right],
\end{equation}
with arbitrary exchange couplings and spin raising and lowering operators at site $j$ defined as $S_j^{\pm} \equiv S_j^{x}\pm iS_j^{y}$.
The Zeeman term for an uni-axial (but not necessarily homogeneous) field is given by
\begin{equation}\label{H_Z}
     H_Z = \sum_{j}h_jS_j^z.
\end{equation}

The starting point of the pm-fRG\cite{Niggemann2021,Niggemann2022,sbierski2023,Schneider2024_Tflow} approach developed for Heisenberg- and XXZ-Hamiltonians is to rewrite the spin-$1/2$ operators using the $SO(3)$ pseudo-Majorana representation\cite{Tsvelik1992,Martin1959}. This employs three auxiliary Majorana operators per site, $\eta_j^{\alpha} \; (\alpha = x,y,z)$, 
\begin{subequations} \label{eq_pm_spin_operators}
\begin{align}
    S_j^z&=-i \eta_j^x \eta_j^y,\\
    S_j^{\pm} &= \left(-i\eta_j^y\mp\eta_j^x\right)\eta_j^z.
\end{align}
\end{subequations} 
The fermionic Majorana operators fulfill $(\eta_j^\alpha)^\dagger = \eta_j^{\alpha}$, canonical anti-commutation relations $\{\eta^{\alpha}_i,\eta^{\beta}_j\}=\delta^{\alpha\beta}\delta_{ij}$ and are normalized as $\left(\eta_j^{\alpha}\right)^2 = 1/2$. 

The presence of a finite $Z$-field as in Eq.~\eqref{H_Z} leads to a mixing in flavor space in the non-interacting fermionic Green's function. This complication has so far hindered the application of the pm-fRG calculations in this case. The goal of this work is to introduce a symmetry-adapted variant of the pm-fRG which keeps the complexity of the flow equations at a level tractable for numerical solution.

%%%%%%%%%%%%%%%%%%%%%%%%%%%%%%%%%%%%%%%%%%%%%%%%%%%%%%%%%%%%%%%%%%%%%%%%%%%%%%%%%
\subsection{Mixed representation}
\label{Symmetry_Adapted_Represenation}

To simplify calculations including magnetic fields, we diagonalize the non-interacting pseudo-Majorana Hamiltonian by combining $\eta^x_j$ and $\eta^y_j$ into a \emph{complex} fermion with creation and annihilation operators $c^\dagger_j,c_j$ as follows\cite{Girvin2019} 
\begin{subequations} \label{eq_eta_x_and_eta_y} 
\begin{align} 
	\eta_j^{x} &= \frac{i}{\sqrt{2}}\left(c_j-c_j^{\dagger}\right) ,\\ 
    \eta_j^{y} &= \frac{1}{\sqrt{2}}\left(c_j+c_j^{\dagger}\right) .
\end{align}
\end{subequations} 
The anti-commutation relations of the complex fermions are $\{c_j^{\dagger},c_{j^\prime}\} = \delta_{j,j^\prime}$. We drop the $z$ superscript of the remaining Majorana fermion, $\eta_j^z \equiv \eta_j$ and note that $	\{c_j^{(\dagger)},\eta_{j^\prime}\} = 0$. With this transformation, the spin operators in Eq.~\eqref{eq_pm_spin_operators} become
\begin{subequations} \label{eq:S_mixed}
\begin{align}
	S^+_j &= -i\sqrt{2}c_j\eta_j ,\\
    S^-_j &= -i\sqrt{2}c_j^{\dagger}\eta_j ,\\
    S_j^z &= \frac{1}{2} - c_j^{\dagger}c_j.
\end{align}
\end{subequations}
This mixed representation which involves both a complex and a Majorana fermion has first been introduced as ``drone-fermion representation'' in Refs.~\onlinecite{Cottam1970,mattis1965,spencer1968}. We note its similarity to the Jordan-Wigner representation \cite{Jordan1928}. However, instead of string operators it is the Majorana operator which ensures the correct commutation relations between spin operators at different sites. 

In the mixed representation \eqref{eq:S_mixed}, the XXZ--Z model \eqref{eq_XXZ--Z} reads
\begin{equation}\label{eq_XXZ}
\begin{split}
	H_{XXZ} & = \sum_{i<j}\left[J_{ij}^{\perp}\left(c_j^{\dagger}c_i\eta_j\eta_i + c_i^{\dagger}c_j\eta_i\eta_j \right) \right. \\
    & -\left. J_{ij}^z\left(c_i^{\dagger}c_j^{\dagger}c_ic_j +\frac{1}{2}c_i^{\dagger}c_i +\frac{1}{2}c_j^{\dagger}c_j -\frac{1}{4}\right) \right],
\end{split}
\end{equation}
\begin{equation}\label{eq_H_Z_def}
	H_Z = \sum_{j}h_j\left(\frac{1}{2} - c_j^{\dagger}c_j \right).
\end{equation} 
Now the non-interacting part \eqref{eq_H_Z_def}  is diagonal in the fermionic flavor, the magnetic field acts as an on-site potential for the $c$-fermions. This will greatly reduce the effort to include the magnetic Z-field in the pm-fRG framework.

%%%%%%%%%%%%%%%%%%%%%%%%%%%%%%%%%%%%%%%%%%%%%%%%%%%%%%%%%%
\subsection{Symmetries}\label{symmetries}
In the following, we discuss the gauge symmetries of the mixed representation \eqref{eq:S_mixed} (which leave the spin operators invariant) and those associated to the XXZ--Z model (which leave the Hamiltonian $H$ invariant). Together, they restrict the set of possible non-zero correlation and vertex functions and thus ultimately simplify the fRG flow equations. Spatial symmetries of the underlying lattice are treated as discussed in Ref.~\onlinecite{Mueller2023}.

\subsubsection*{Local $\mathbb{Z}_2$ gauge symmetry}
The spin operators in Eq.~\eqref{eq:S_mixed} are invariant under a local $\mathbb{Z}_2$ gauge transformation that acts on all fermionic operators at site $j$ as $\Phi_j \equiv ( c_j, c_j^{\dagger}, \eta_j )^{\rm{T}} \rightarrow -\Phi_j$.
This enforces bi-local correlations which means that all correlators with an odd number of fermionic site-$j$ operators vanish.
The $\mathbb{Z}_2$ symmetry is associated with an artificial degeneracy of the pseudo-fermionic eigenstates, with a degeneracy factor of $2^{N/2}$ for $N$ lattice sites \cite{Tsvelik1992} reflecting all possible fermion parities for pairs of sites.

\subsubsection*{Anti-unitary symmetry}
The spin operators in Eq.~\eqref{eq:S_mixed} are further invariant under a global anti-unitary symmetry 
\begin{equation} 
au:\;i\rightarrow-i, \; \eta_j \rightarrow-\eta_j \; \forall j. \label{eq_sym_au}
\end{equation}
This symmetry forces real-valued correlators with an odd number of Majorana operators to vanish. Moreover, it also relates frequencies $\pm\omega$ in Fourier-transformed imaginary time ordered correlators to be defined below. This is further discussed in Appendix \ref{Freq_syms}.

We now turn to the specific symmetries of the Hamiltonian operators, \eqref{eq_XXZ} and \eqref{eq_H_Z_def}.

\subsubsection*{Global $U(1)$ Symmetry}
Both parts of the XXZ--Z Hamiltonian \eqref{eq_XXZ--Z} are invariant under a global $U(1)$ spin rotation symmetry around the spin-$Z$ axis. In the mixed representation this amounts to 
\begin{equation}\label{U1}
 U(1):\;\;  c_j\rightarrow e^{i\theta}c_j,\;c_j^{\dagger}\rightarrow e^{-i\theta}c_j^{\dagger}\; \forall j,
\end{equation}
and corresponds to $c$-fermion conservation.
If the $U(1)$ symmetry is unbroken, only correlators including $c,c^{\dagger}$ in a pairwise fashion can exist because
\begin{equation}
    \gsexpval{c^{\dagger}c^{\dagger}} \overset{U(1)}{=} e^{-2i\theta}\gsexpval{c^{\dagger}c^{\dagger}} \forall \theta \Rightarrow \gsexpval{c^{\dagger}c^{\dagger}} = 0.
\end{equation}

\subsubsection*{$h_j=0$: Time-reversal and particle-hole symmetry}
Without magnetic fields $h_j = 0$, the XXZ-Hamiltonian is symmetric under time-reversal $S_j^\alpha \rightarrow -S_j^\alpha \; \forall j,\alpha$ which is anti-unitary and translates to the mixed representation \eqref{eq:S_mixed} as follows 
 \begin{equation} 
 TR:\;\; i\rightarrow-i,\;c\rightarrow c^{\dagger},\;c^{\dagger}\rightarrow c.
 \end{equation}
Moreover, there exists a global particle hole symmetry
\begin{equation}
PH:\;\; c_j \leftrightarrow c_j^{\dagger}.
\end{equation}
In combination with time-reversal, this ensures that for $h_j=0$ all vertices are purely real.

%%%%%%%%%%%%%%%%%%%%%%%%%%%%%%%%%%%%%%%%%%%%%%%%%%%%%%%%%%%%%%%%%%%%%%%%%%%%%%%%%%%%%
%%%%%%%%%%%%%%%%%%%%%%%%%%%%%%%%%%%%%%%%%%%%%%%%%%%%%%%%%%%%%%%%%%%%%%%%%%%%%%%%%%%%%
\section{Green's functions and action}\label{Greens_function_and_action}

In this section, we discuss the Grassmann action\cite{Altland2006} for the mixed representation and proceed to define the Green's function objects which are at the core of the diagrammatic method and in particular the fRG. We assume that the symmetries from Sec.~\ref{symmetries}, except TR and PH, are unbroken. Since we have performed a unitary transformation of the pseudo-Majorana operators, the action formalism for the $SO(3)$ representation\cite{Nilsson2013} can be applied to derive the two local two-point Green's functions $G_{c_i^{\dagger}c_i}, G_{\eta_i\eta_i}$. 
A fermionic two-point Matsubara Green's function in thermal equilibrium at temperature $T=1/\beta$ is defined as 
\begin{subequations} 
\begin{align}\label{eq_def_greensfunction}
	\left[\mathbf{G} \right]_{12} & = \int_0^{\beta}d\tau_1 d\tau_2 e^{i\omega_1\tau_1+i\omega_2\tau_2} \langle \xi_2(\tau_2)\xi_1(\tau_1) \rangle \nonumber \\
    & \equiv \gsexpval{\xi_2\xi_1} = - \gsexpval{\xi_1\xi_2} = -\left[\mathbf{G} \right]_{21}\\
	 & \equiv - \frac{1}{\beta}	\delta_{j_1 j_2}\delta_{\omega_1,-\omega_2} G_{12}(\omega_1).
\end{align}
\end{subequations} 
Here, $\xi_{1}$ (and $\xi_{2}$) are elements of the Grassmann super-field vector $\Xi_1 = \left(\psi_1, \bar{\psi}_1, \zeta_1 \right)^{\rm{T}}$ related to the coherent states of $c,c^{\dagger}$ and $\eta$ respectively\cite{Altland2006}. The multi-index $1$ contains all information on particle type, lattice site $j_1$ and Matsubara frequency $\omega_1$ (short for $\omega_{n_1}=[2n_1+1]\pi T, \;n_1\in \mathbb{Z} $) which appear in the Fourier transformation
\begin{equation}
    \xi_1\left(\tau\right) \equiv \frac{1}{\beta}\sum_{\omega_1} e^{-i\omega_{1}\tau} \xi_1\left(\omega_{1}\right).
\end{equation}
Note that unlike the standard convention for complex fermions, we use the same sign in the exponent regardless of particle type following Ref.~\onlinecite{Niggemann2021}.

In the Grassmann field formalism, we write the action as\cite{Kopitz2010}
\begin{widetext}
\begin{subequations} \label{eq_action_XXZ}
	\begin{align}
	S[\Xi]
			& = S_0[\Xi] + S_{I}[\Xi] \\
    S_0[\Xi] & = 
	\frac{1}{\beta} \sum_{j} \sum_{\omega_1,\omega_2} \left[  \frac{1}{2} \Xi^T_j(\omega_1) i\omega_2 \Xi_j(\omega_2) \right. 
    -  h_j\  \bar{\psi}_j(\omega_1)\psi_j (\omega_2)  \Bigr] \delta_{\omega_1,-\omega_2} \equiv  \frac{1}{2\beta^2} \sum_{1,2} \Xi^T_1 \left[\mathbf{G}_0^{-1} \right]_{12} \Xi_2 \label{eq_action_G0} \\
	S_{I}[\Xi] & = -\frac{1}{2\beta} \sum_{i<j} \sum_{\omega_1,\omega_2}   J_{ij}^{z}\Bigl(\bar{\psi}_i(\omega_1)\psi_i(\omega_2) + \bar{\psi}_j(\omega_1)\psi_j (\omega_2)  \Bigr) 
  \delta_{\omega_1,-\omega_2} \nonumber\\
		 &+
  \frac{1}{\beta^3} \sum_{i<j} \sum_{\omega_1,\omega_2,\omega_3,\omega_4} \biggl[ J_{ij}^{\perp}\Bigl(\bar{\psi}_j(\omega_1)\psi_i(\omega_2)\zeta_j(\omega_3)\zeta_i (\omega_4) 
    +  \bar{\psi}_i(\omega_1)\psi_j(\omega_2)\zeta_i(\omega_3)\zeta_j (\omega_4)\Bigr)  \nonumber\\
		& -
   J_{ij}^z \bar{\psi}_i(\omega_1)\bar{\psi}_j(\omega_2)\psi_i(\omega_3)\psi_j(\omega_4)  \biggr]
\delta_{ \omega_1 + \omega_2, - \omega_3 - \omega_4 } 
	%\beta \sum_{ij} \left[  \frac{1}{2} \delta_{ij} h_i + \frac{1}{4} J_{ij}^z  \right].
\end{align}
\end{subequations}
\end{widetext}
Here and in the following, constant terms have been neglected. Note that the first line of $S_I[\Xi]$, which is only bi-linear in the fields and thus non-interacting, will cancel against the Hartree-like contribution from the last term (see Appendix \ref{intialconditions}). Using the Dyson equation $\left[\mathbf{G}\right]=\left[\mathbf{G}_{0}^{-1}-\boldsymbol{\Sigma}\right]^{-1}$ and the choice of $\mathbf{G}_{0}$ in Eq.~\eqref{eq_action_G0}, we obtain for the full Green's functions (see Appendix \ref{Appendix_G0} for details)
\begin{subequations} \label{eq_G_matrix}
\begin{align}
	\left[\mathbf{G} \right]_{\psi_1\bar{\psi}_2} &=  - \frac{1}{\beta} \delta_{j_1j_2}\delta_{\omega_1,-\omega_2} \frac{1}{i\omega_1 - h_{j_1} + \Sigma_{\bar{\psi}_2\psi_1}\left(\omega_1\right) } \nonumber \\
    &\equiv - \frac{1}{\beta} 	\delta_{j_1j_2}\delta_{\omega_1,-\omega_2} G_{\psi_1\bar{\psi}_2}(\omega_1), \\
	\left[\mathbf{G}\right]_{\bar{\psi}_1\psi_2} &= - \frac{1}{\beta} \delta_{j_1j_2}\delta_{\omega_1,-\omega_2} \frac{1}{i\omega_1 +h_{j_1} + \Sigma_{\psi_2\bar{\psi}_1}\left(\omega_1\right)} \nonumber \\
    &\equiv - \frac{1}{\beta} \delta_{j_1j_2}\delta_{\omega_1,-\omega_2}G_{\bar{\psi}_1\psi_2}(\omega_1), \\
	\left[\mathbf{G}\right]_{\zeta_1\zeta_2} &=  - \frac{1}{\beta} \delta_{j_1j_2}\delta_{\omega_1,-\omega_2}  \frac{1}{ i\omega_1 + \Sigma_{\zeta_2\zeta_1}\left(\omega_1\right)} \nonumber \\
    &\equiv - \frac{1}{\beta} \delta_{j_1j_2}\delta_{\omega_1,-\omega_2} G_{\zeta_1\zeta_2}(\omega_1).
\end{align}
\end{subequations} 
From the anti-symmetry of $\left[\mathbf{G} \right]_{\psi_1\bar{\psi}_2}$, see Eq.~\eqref{eq_def_greensfunction}, it follows $
	 G_{\bar{\psi}_1\psi_2}(\omega_1) = - G_{\psi_2\bar{\psi}_1}(-\omega_1)$ and hence also the anti-symmetry of the self-energies
\begin{equation}
	 \Sigma_{\psi_2\bar{\psi}_1}\left(\omega_1\right) =- \Sigma_{\bar{\psi}_1\psi_2}\left(-\omega_1\right) .
\end{equation}
Here we used the following definitions
\begin{subequations} 
\begin{align}\label{eq_def_sigma}
	\left[\mathbf{\Sigma} \right]_{\psi_1\bar{\psi}_2} & \equiv \beta \delta_{j_1j_2}\delta_{\omega_1,-\omega_2} \Sigma_{\psi_1\bar{\psi}_2}(\omega_2), \\
	\left[\mathbf{\Sigma}\right]_{\bar{\psi}_1\psi_2} & \equiv \beta \delta_{j_1j_2}\delta_{\omega_1,-\omega_2}\Sigma_{\bar{\psi}_1\psi_2}(\omega_2), \\
	\left[\mathbf{\Sigma}\right]_{\zeta_1\zeta_2} & \equiv \beta \delta_{j_1j_2}\delta_{\omega_1,-\omega_2} \Sigma_{\zeta_1\zeta_2}(\omega_2).
\end{align}
\end{subequations} 

Beyond the above fermionic two-point correlators, we also consider fermionic four-point correlators defined as
\begin{equation}
\begin{split}
    G_{1234} \equiv  & \int_0^{\beta}  d\tau_1 d\tau_2 d\tau_3 d\tau_4 \Bigl( e^{i\omega_1\tau_1+i\omega_2\tau_2+i\omega_3\tau_3+i\omega_4\tau_4} \\
    & \cdot \langle \xi_4(\tau_4)\xi_3(\tau_3)\xi_2(\tau_2)\xi_1(\tau_1) \rangle \Bigr).
\end{split}
\end{equation}
The one-particle irreducible contributions to $G_{1234}$ are the four-point vertices $\Gamma_{1234}$, the formal connection is given via the tree expansion\cite{Kopitz2010}. Out of the $3^4=81$ possible combinations of field-types, most vanish due to $U(1)$ and $\mathbb{Z}_2$ symmetry. Using the anti-symmetry under multi-index exchange and the bi-local nature only five different four-point vertices are independent, 
$\Gamma_{\bar{\psi}_i\bar{\psi}_i\psi_j\psi_j},\ \Gamma_{\bar{\psi}_i\bar{\psi}_j\psi_i\psi_j},\ \Gamma_{\bar{\psi}_i\psi_i\zeta_j\zeta_j},\ \Gamma_{\bar{\psi}_i\psi_j\zeta_i\zeta_j}$ and $\Gamma_{\zeta_i\zeta_i\zeta_j\zeta_j}$. Moreover, energy conservation is taken into account via reduction to three bosonic transfer frequencies\cite{Niggemann2021}  
\begin{equation}
	\Gamma_{1234}(s,t,u) \equiv \beta \delta_{0,\omega_1 + \omega_2 + \omega_3+ \omega_4} \Gamma_{1234}(\omega_1,\omega_2,\omega_3,\omega_4),
\end{equation} where we define
\begin{equation}\label{eq_transferfrequencies}
	\begin{aligned}
		s &\equiv \omega_1 + \omega_2 &= -\omega_3 -\omega_4 ,\\
		t &\equiv \omega_1 + \omega_3 &= -\omega_2 - \omega_4 ,\\
		u &\equiv \omega_1 + \omega_4 &= -\omega_2 - \omega_3 .\\
	\end{aligned}
\end{equation}

\begin{figure}[H]
	\centering
	\includegraphics[scale=1]{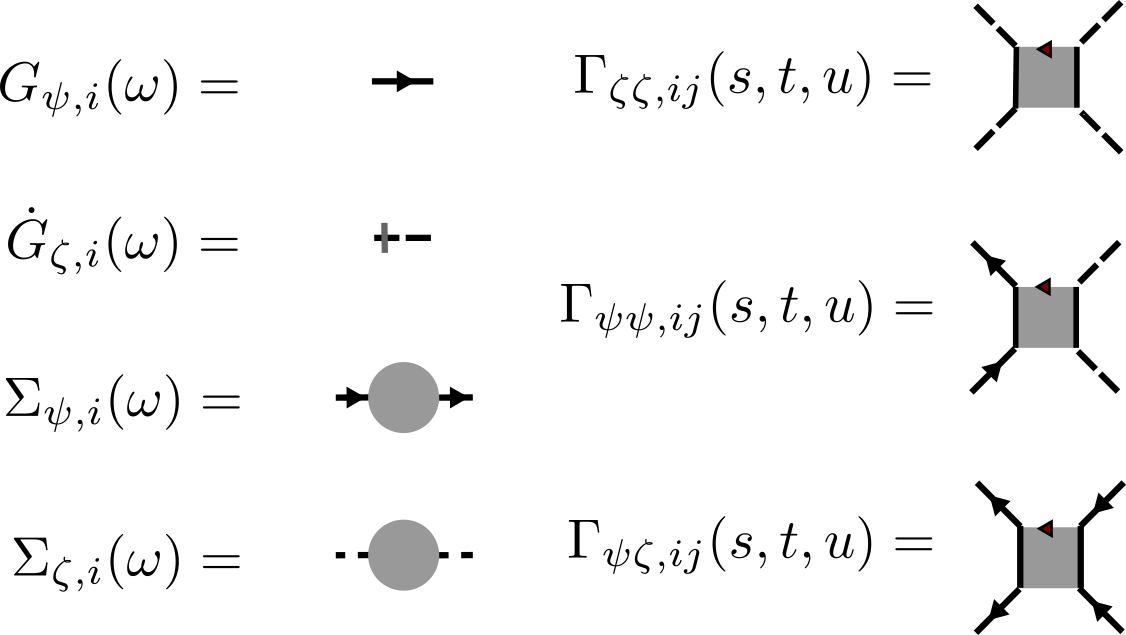}
	\caption{Diagrammatic representation of the two- and four-point objects defined in Eqns.~\eqref{eq_Sigma_final}-\eqref{G_4_def}. Dashed lines correspond to the field $\zeta$ representing Majorana operator $\eta$, full lines with arrows pointing away (towards) the vertex correspond to $\bar{\psi}$ ($\psi$) representing the complex fermion. The small arrow at the box boundary points towards the first index of $\Gamma_{1234}$, the others follow in an anti-clockwise sense.}
	\label{fig_Grafics_guide}
\end{figure}

For a lighter notation, we finally drop all redundant labels. For the two-point objects we write
\begin{subequations} \label{eq_Sigma_final}
\begin{align}
	\Sigma_{\psi,j}(\omega) & \equiv \Sigma_{\bar{\psi}_j\psi_j}(\omega), \\
	\Sigma_{\zeta,j}(\omega) &\equiv \Sigma_{\zeta_j\zeta_j}(\omega), 
\end{align}
\end{subequations} 
with $\omega$ on the left-hand side referring to the \emph{second} field on the right-hand side, cf.~Eq.~\eqref{eq_def_sigma}. Equivalently for the full propagators, with $\omega$ referring to the \emph{first} field's frequency [cf.~Eq.~\eqref{eq_def_greensfunction}], we define
\begin{subequations} \label{eq_G_final}
\begin{align}
	G_{\psi,j}(\omega) & \equiv G_{\bar{\psi}_j\psi_j}(\omega), \\
	G_{\zeta,j}(\omega) & \equiv G_{\zeta_j\zeta_j}(\omega). 
\end{align}
\end{subequations} 
For the four-point vertices we simplify the notation to account explicitly for their bi-local nature, 
\begin{subequations} \label{G_4_def}
\begin{align}
	\Gamma_{\psi\psi,ij}\left(s,t,u\right) & \equiv \Gamma_{\bar{\psi}_i\bar{\psi}_i\psi_j\psi_j}\left(s,t,u\right),  
	\\
    \Gamma^{\times}_{\psi\psi,ij}\left(s,t,u\right) & \equiv \Gamma_{\bar{\psi}_i\bar{\psi}_j\psi_i\psi_j}\left(s,t,u\right), \\
	\Gamma_{\psi \zeta,ij}\left(s,t,u\right) & \equiv \Gamma_{\bar{\psi}_i\psi_i\zeta_j\zeta_j}\left(s,t,u\right),  
	\\
    \Gamma^{\times}_{\psi \zeta,ij}\left(s,t,u\right) & \equiv \Gamma_{\bar{\psi}_i\psi_j\zeta_i\zeta_j}\left(s,t,u\right), \\
	\Gamma_{\zeta\zeta,ij}\left(s,t,u\right) & \equiv \Gamma_{\zeta_i\zeta_i\zeta_j\zeta_j}\left(s,t,u\right). 
\end{align}
\end{subequations} 
The superscript $\times$ indicates that the \emph{first} and \emph{third} field are site-paired, without the superscript the pairing is between the \emph{first} and \emph{second} field. Furthermore, the anti-symmetry under exchange of two fields can be used to restrict the required frequency combinations, see Appendix~\ref{Freq_syms}. The diagrammatic representation of these objects is given in Fig.~\ref{fig_Grafics_guide}.

%%%%%%%%%%%%%%%%%%%%%%%%%%%%%%%%%%%%%%%%%%%%%%%%%%%%%%%%%%%%%%%%%%%%%%%%%%%%%%%%%%%%%%
%%%%%%%%%%%%%%%%%%%%%%%%%%%%%%%%%%%%%%%%%%%%%%%%%%%%%%%%%%%%%%%%%%%%%%%%%%%%%%%%%%%%%%
\section{Observables in the mixed representation}
\label{sec_observables}

In this section, we discuss spin observables and their expressions in terms of the two- and four-point functions of the pseudo-fermionic mixed representations. We focus on the magnetization in $Z$-direction and the static spin susceptibility. As in both cases we are calculating operator averages, $\gsexpval{O} = \tr \left(O\rho\right) / \tr\left(\rho\right)$, the aforementioned $2^{N/2}$-fold degeneracy of the spin representation reflected in the parity-sectors of the density matrix $\rho \sim e^{-\beta H}$ is simply canceled between numerator and denominator\cite{Niggemann2021}.

%%%%%%%%%%%%%%%%%%%%%%%%%%%%%%%%%%%%%%%%%%%%%%%%%%%%%%%%%%%%%%%%%%%%%%%%%%%%%%%%%%%%
\subsection{Magnetization}\label{Magnetization}
One main objective of this work is to compute the magnetization in $Z$-direction which is related to the $c-$fermion two-point function. For site $j$, we have from Eq.~\eqref{eq_pm_spin_operators},
\begin{equation}\label{M_eq_first_step}
	\begin{split}
		M_j & \equiv \gsexpval{S_j^z} = \frac{1}{2} - \lim_{\epsilon\rightarrow0} \gsexpval{ \bar{\psi}_j(\tau+\epsilon)\psi_j(\tau) }  \\
		& = 
		\frac{1}{2} - \lim_{\epsilon\rightarrow0} \frac{1}{\beta}\sum_{\omega} e^{i\omega\epsilon} \gsexpval{ \bar{\psi}_j(\omega)\psi_j(-\omega) } \\
		& =
		\frac{1}{2} - \lim_{\epsilon\rightarrow0} \frac{1}{\beta} \sum_{\omega} e^{i\omega\epsilon} G_{\psi,j}(\omega) , \\
	\end{split}
\end{equation}
%Note that Kopietz's convention for correlators yields a minus sign from the definition of the Green's function. 
where the positive infinitesimal $\epsilon$ ensures the proper operator ordering. Numerically, such sums over Green's functions converge very poorly\cite{FB2024} and part of the sum should be performed analytically. The convergence generating factor $e^{i\omega\epsilon}$ with $\epsilon \geq 0, \  \epsilon\rightarrow 0$ is omitted in the following. We use Eq.~\eqref{eq_G_matrix} and split into real and imaginary part, 
\begin{equation}
	\begin{split}
		\sum_{\omega} &G_{\psi,j}(\omega) = \sum_{\omega} \frac{1}{i\omega + h_j -\Sigma_{\psi,j}(-\omega)} \\
		& = 
		\sum_{\omega} \frac{ h_j - \textrm{Re}\left[\Sigma_{\psi,j}(\omega)\right] }{\left(h_j-\textrm{Re}\left[\Sigma_{\psi,j}(\omega)\right]\right)^2 + \left(\omega + \textrm{Im}\left[\Sigma_{\psi,j}(\omega) \right] \right)^2} \\
		& + i \sum_{\omega} \frac{ \omega + \textrm{Im}\left[\Sigma_{\psi,j}(\omega)\right] }{\left(h_j-\textrm{Re}\left[\Sigma_{\psi,j}(\omega)\right]\right)^2 + \left(\omega + \textrm{Im}\left[\Sigma_{\psi,j}(\omega)\right]  \right)^2} .
	\end{split}
\end{equation}
The imaginary part is asymmetric in $\omega$ and only the large-$\omega$ tail contributes, for which we re-establish the convergence factor. From
\begin{equation}
	\lim_{\epsilon\rightarrow0} i \sum_{\omega} e^{i\omega\epsilon} \frac{1}{\omega} = \frac{\beta}{2}.
\end{equation}
we observe that this contribution cancels the $1/2$ in Eq.~\eqref{M_eq_first_step}. In conclusion,
\begin{equation}
	M_j = \frac{1}{\beta} \sum_{\omega} \frac{\textrm{Re}\left[\Sigma_{\psi,j}(\omega)\right] - h_j}{\left(h_j-\textrm{Re}\left[\Sigma_{\psi,j}(\omega)\right]\right)^2 + \left(\omega + \textrm{Im}\left[\Sigma_{\psi,j}(\omega) \right] \right)^2}.
\end{equation}
In practice it is advisable to extrapolate $\Sigma_{\psi,j}(\omega)$ beyond the numerically computed frequency box, see Appendix \ref{Freq_syms}.

%%%%%%%%%%%%%%%%%%%%%%%%%%%%%%%%%%%%%%%%%%%%%%%%%%%%%%%%%%%%
\subsection{Spin susceptibilities}

The static spin susceptibility defined as $\chi_{ij}^{zz} = - \del M_i / \del h_j$ can also be computed from the four-point fermionic correlator. Generalizing to arbitrary directions $\alpha,\alpha^\prime$ in spin space, the definition reads
\begin{equation}\label{eq_suceptibilitites_observable}
\begin{split}
	\chi&_{ij}^{\alpha\alpha'} = \int_{0}^{\beta} d\tau \gsexpval{S_i^{\alpha}(\tau)S_j^{\alpha'}(0)}_c \\
    &= \int_{0}^{\beta} d\tau \gsexpval{\Bigl(S_i^{\alpha}(\tau)- \gsexpval{S_i^{\alpha}(\tau)}\Bigr)\left(S_j^{\alpha'}(0)- \gsexpval{S_i^{\alpha^{\prime}}(0)}\right)},
\end{split}
\end{equation}
but only the cases $\alpha = \alpha' = z$ and $\alpha = \alpha' = x$ [equal to $\alpha = \alpha' =y$ by $U(1)$ symmetry] are non-zero and independent. Inserting the mixed spin representation \eqref{eq:S_mixed} and relating the four-point function to the vertices $\Gamma$ via the tree expansion\cite{Kopitz2010}, we arrive at
\begin{equation}\label{eq_suceptibility_zz}
	\begin{split}
		\chi_{ij}^{zz}
		& = 
		\frac{1}{\beta^2}\sum_{\omega, \omega'} \left[ G_{\psi,i}^2(\omega) G_{\psi,j}^2(\omega')\right.\\ 
        &\cdot \Gamma_{\psi\psi,ij}^{\times}\left(\omega+\omega',0,\omega-\omega'\right) 
		 \left. - \beta \delta_{ij}\delta_{\omega,\omega'} G_{\psi,i}^2(\omega)\right]. 
	\end{split}
\end{equation}
and
\begin{equation}
	\begin{split}
		\chi_{ij}^{xx}
		& =
		\frac{1}{\beta^2} 	 \sum_{\omega\omega'}\left[-\beta\delta_{ij}\delta_{\omega,\omega'} G_{\psi,i}(\omega)G_{\zeta,i}(\omega) \right. \\
		&+ G_{\psi,j}(\omega) G_{\psi,i}(-\omega') G_{\zeta,j}(\omega) G_{\zeta,i}(\omega') \\
        &\cdot \left. \Gamma_{\psi\zeta,ij}^{\times}(-\omega-\omega',0,\omega-\omega') \right]. \\
	\end{split}
\end{equation}
If needed, it is straightforward to also extract dynamic spin susceptibilities at non-zero bosonic frequency $i\Omega_n$, see Ref.~\onlinecite{Mueller2023}. Then a sum over this frequency would yield the equal-time spin correlator. Although not required in the following, we also note the important role of the spin susceptibilities in a finite-size scaling approach to detect magnetic phase transitions, see e.g.~Refs.~\onlinecite{Niggemann2022,schneider2024tamingspins}.

%%%%%%%%%%%%%%%%%%%%%%%%%%%%%%%%%%%%%%%%%%%%%%%%%%%%%%%%%%%%%%%%%%%%%%%%%%%%%%%%%%%%%%
%%%%%%%%%%%%%%%%%%%%%%%%%%%%%%%%%%%%%%%%%%%%%%%%%%%%%%%%%%%%%%%%%%%%%%%%%%%%%%%%%%%%%%
\section{FRG Flow Equations in the Mixed Representation}
\label{flow_equations_m_rep}

The basic idea of the fRG\cite{WETTERICH1993,Kopitz2010,metznerFunctionalRenormalization2012} is to modify the bare propagator  $\mathbf{G}_{0}$ in Eq.~\eqref{eq_action_G0} by a deformation (or cut-off) quantified the scalar $\Lambda$. Variation of $\Lambda$ defines a flow. By convention, the starting point is $\Lambda=\infty$ where the action is trivial such that all one-particle irreducible vertices are known exactly. Then a hierarchy of flow-equations is derived that govern the variation of the vertices with $\Lambda$ and in particular allow to compute them in the case $\Lambda=0$ where we recover the undeformed physical action. However, truncations are necessary to solve the flow equations in practice. In the following, we build on the general super-field formulation of fRG put forward in Ref.~\onlinecite{Kopitz2010}.

As the main result of this work we here present the fRG flow equations for the XXZ-Z model in the mixed representation.
In general, the form of the fRG flow equations are independent of the chosen cut-off, in this work however, we apply a multiplicative Lorentzian cut-off $\mathbf{G}_0^\Lambda = \theta_\Lambda \mathbf{G}_0$ in frequency further detailed in Sec.~\ref{sec_cut_off}. We proceed with the flow equations of the self-energy and discuss the overall structure of the flow equations for the four-point vertices using one example. The full flow equations of the four-point vertices are given in Appendix \ref{Appendix_Flow_Equations}. There we also introduce the Katanin truncation\cite{Katanin2004} which we use to include parts of the six-point vertex. The initial conditions for the flow are derived in Appendix \ref{intialconditions}.

%%%%%%%%%%%%%%%%%%%%%%%%%%%%%%%%%%%%%%%%%%%%%%%%%%%%%%%%%%%%%
\subsection{Cut-off and single-scale propagator}\label{sec_cut_off}

In this work, we follow Refs.~\onlinecite{Niggemann2021,Niggemann2022,sbierski2023} and apply a Lorentzian cut-off function to smoothly switch off the bare propagator at Matsubara frequencies smaller in magnitude than $\Lambda$,
\begin{align}
	\theta_{\Lambda}(\omega) &= \frac{\omega^2}{\omega^2+\Lambda^2},&
	\frac{\partial_{\Lambda}\theta_{\Lambda}}{\theta^2_{\Lambda}} &= -\frac{2\Lambda}{\omega^2}.
\end{align}
The flow equations depend on the single-scale propagators\cite{Kopitz2010}, $\dot{\mathbf{G}}^{\Lambda}  \equiv - \mathbf{G}^{\Lambda} (\partial_{\Lambda}{\mathbf{G}^{\Lambda}_0}^{-1})\mathbf{G}^{\Lambda}$, given by
\begin{subequations} 
\begin{align}\label{eq_G_dot_matrix}
	\left[\dot{\mathbf{G}}^{\Lambda}\right]_{\bar{\psi}_1\psi_2} 
    &= - \frac{\partial_{\Lambda}\theta_{\Lambda}}{\beta\theta^2_{\Lambda}}  \frac{\delta_{j_1 j_2} \delta_{\omega_1,-\omega_2}\left( i\omega_1 - h_{j_1} \right) }{ \left( \left(i\omega_1 +h_{j_1}\right)\theta^{-1}_{\Lambda}(\omega_1)+ \Sigma_{\psi_2\bar{\psi}_1}\left(\omega_1\right) \right)^2}   \nonumber \\ 
		& \equiv - \frac{1}{\beta} \delta_{j_1j_2}\delta_{\omega_1,-\omega_2}\dot{G}^{\Lambda}_{\psi,j_1}(\omega_1) , \\
	\left[\dot{\mathbf{G}}^{\Lambda}\right]_{\zeta_1\zeta_2} &= - \frac{\partial_{\Lambda}\theta_{\Lambda}}{\beta\theta^2_{\Lambda}} \frac{\delta_{j_1 j_2}\delta_{\omega_1,-\omega_2} i\omega_1 }{ \left(i\omega_1\theta^{-1}_{\Lambda}(\omega_1)  + \Sigma_{\zeta_2\zeta_1}\left(\omega_1\right)\right)^2 }    \nonumber \\  &
		\equiv -  \frac{1}{\beta} \delta_{j_1 j_2}\delta_{\omega_1,-\omega_2}\dot{G}^{\Lambda}_{\zeta,j_1}(\omega_1).
\end{align}    
\end{subequations} 
Here the abbreviated notation is analogous to Eq.~\eqref{eq_G_final} for the Green's functions.

%%%%%%%%%%%%%%%%%%%%%%%%%%%%%%%%%%%%%%%%%%%%%%%%%
\subsection{Self-energy flow}\label{sec_2_flow}

\begin{figure}[H]
	\centering
	\includegraphics[scale=1]{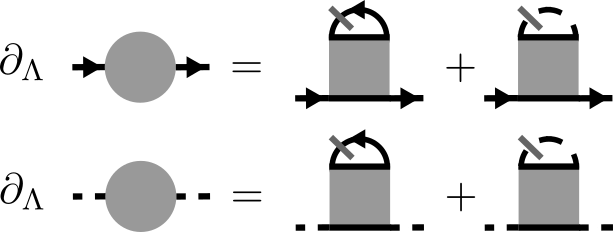}
	\caption{Diagrammatic representation of $\del_{\Lambda}\Sigma^{\Lambda}_{\zeta,i}$ \eqref{eq_2flow_m_f} and $\del_{\Lambda}\Sigma^{\Lambda}_{\psi,i}$ \eqref{eq_2flow_c_f}. According to Fig.~\ref{fig_Grafics_guide}, connected lines represent fields on the same lattice site. A closed loop corresponds to a sum over internal site and frequency indices.}
	\label{fig_dia_2_mm_cc}
\end{figure}

The flow equations for the fermionic self-energy are
\begin{subequations} 
\begin{equation} \label{eq_2flow_c_f}
	\begin{split}
		\del_{\Lambda}\Sigma&_{\psi,i}^{\Lambda}(-\omega)  
		= 
		\frac{1}{2\beta} \sum_{j} \sum_{\omega'}\\
        &\left[ -2 \dot{G}^{\Lambda}_{\psi,j}\left(\omega' \right) \Gamma^{\Lambda\times}_{\psi\psi,ji}\left( \omega'+\omega,0, \omega'-\omega \right) \right.\\
		& \quad \left. +  
		\dot{G}^{\Lambda}_{\zeta,j}\left(\omega' \right) \Gamma^{\Lambda}_{\psi\zeta,ij}\left(  0, \omega'+\omega, -\omega'+\omega \right)  \right],
	\end{split}
\end{equation}
\begin{equation} \label{eq_2flow_m_f}
	\begin{split}
		\del_{\Lambda}&\Sigma_{\zeta,i}^{\Lambda}(-\omega)
		= 
		\frac{1}{2\beta} \sum_{j} \sum_{\omega'} \\
        & \left[ 2 \dot{G}^{\Lambda}_{\psi,j}\left(\omega'\right) \Gamma^{\Lambda}_{\psi\zeta,ji}\left(  0, \omega'+\omega, \omega'-\omega \right) \right.\\
		& \quad \left. +  
		\dot{G}^{\Lambda}_{\zeta,j}\left(\omega' \right) \Gamma^{\Lambda}_{\zeta\zeta,ji}\left(  0, \omega'+\omega, \omega'-\omega \right)  \right].
	\end{split}
\end{equation}
\end{subequations} 
The corresponding diagrams are shown in Fig.~\ref{fig_dia_2_mm_cc}. Note that the mixed four-point vertices involving the $\psi$ and $\zeta$ fields appear on the right-hand side of these flow equations. This is also the case for the flow of the four-point vertices discussed next. 

%%%%%%%%%%%%%%%%%%%%%%%%%%%%%%%%%%%%%%%%%%%%%%%%%%
\subsection{Flow equations for four-point vertex}
\label{sec_4_flow}

We finally turn to the flow equations for the four-point vertices $\Gamma^\Lambda$ where the six-point vertex is approximated to remain at its initial condition which is zero. [However, we involve parts of the six-point vertex using the Katanin truncation, see Appendix \ref{Katanin}]. Here, we restrict ourselves to a discussion of $\partial_{\Lambda}\Gamma^\Lambda_{\zeta\zeta,ij}$ to illustrate the nature of the flow equations, see Fig.~\ref{fig_single_4_vertex}. We provide the full set of flow equations in Appendix~\ref{Appendix_Flow_Equations}. The flow equation involves all possible symmetry-allowed combinations of vertices. This includes combinations of pure Majorana vertices but also the mixed vertices. Furthermore terms can be split into three groups, see the boxes in Fig.~\ref{fig_single_4_vertex}. Terms including internal site sums (third line in Fig.~\ref{fig_single_4_vertex}) depend on an internal propagator loop characterized by the bosonic frequency $s$, while those that have crossed (amputated) external legs are characterized by $t$ and all other terms by $u$, respectively. For pure pm-fRG, terms that depend on $u$ and $t$ are symmetry related by the exchange of external legs. In our formalism however, this symmetry only exist in flow equations for non-crossed vertices $\partial_{\Lambda}\Gamma^\Lambda$ and not for crossed vertices $\partial_{\Lambda}\Gamma^{\Lambda\times}$. If the considered flow-equation is local $i=j$, further simplifications are possible as crossed and non-crossed vertices are equal in that case.
\begin{figure}[H]
	\centering
	\includegraphics[scale=1]{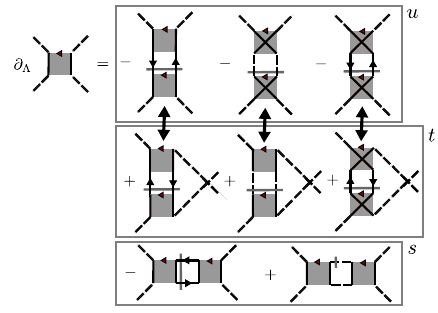}
	\caption{ Diagrammatic representation of $\partial_{\Lambda}\Gamma^\Lambda_{\zeta\zeta,ij}$ grouped into $s,\ t,\ u$ channels characterizing the dependence of the internal propagator loop on the external frequencies. The $t$ and $u$ channel are related by an exchange of external legs. Frequencies and prefactors are omitted here for clarity, they are given in Appendix~\ref{4flowequations}. }
	\label{fig_single_4_vertex}
\end{figure}

%%%%%%%%%%%%%%%%%%%%%%%%%%%%%%%%%%%%%%%%%%%%%
%%%%%%%%%%%%%%%%%%%%%%%%%%%%%%%%%%%%%%%%%%%%%
\section{Benchmark tests}
\label{Benchmarking}
To test the implementation and capabilities of the $U(1)$-pm-fRG we now proceed to benchmark it against exact analytical results and quantum Monte Carlo (QMC). We also checked that the results  applied to various Heisenberg models with $SU(2)$ symmetry reproduce data obtained with the standard pm-fRG\cite{Niggemann2021,Niggemann2022}. In the following we use the Katanin truncation of the flow equations, with the results presented in Section \ref{Dimer_Benchmark} as the sole exception.
Details on the numerical implementation are given in Appendix~\ref{Numerical_Implementation}.
%%%%%%%%%%%%%%%%%%%%%%%%%%%%%%%%%%%%%%%%%%%%%
\subsection{XXZ-dimer}
\label{Dimer_Benchmark}
\begin{figure*}
	\centering
	\includegraphics[scale=1.0]{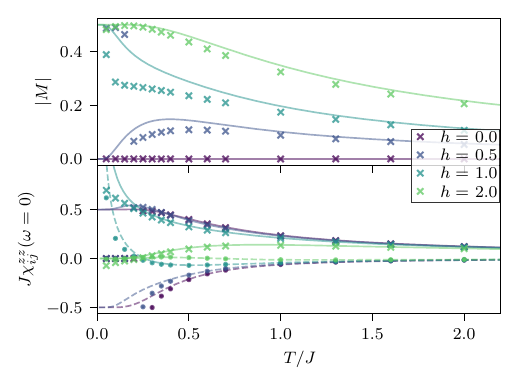}
    \includegraphics[scale=1.0]{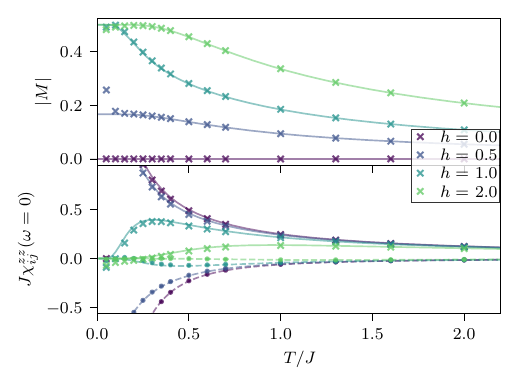}
	\caption{(a) AFM Heisenberg dimer $J^z=1,\ J^{\perp}=1$: Magnetization $|M|$ (top) and static susceptibility $\chi_{ij}^{zz}(i\omega_m = 0)$ (bottom) as a function of temperature for various external magnetic fields $h$. Lines correspond to the exact solution while symbols denote fRG results. For the susceptibility we distinguish the local (crosses) and non-local (dots) case. (b) The same as in (a) but for the Ising case $J^z=1,\ J^{\perp}=0$.}
	\label{AFM_Dimer}
\end{figure*}

\begin{figure}
	\centering
	\includegraphics[scale=1.0]{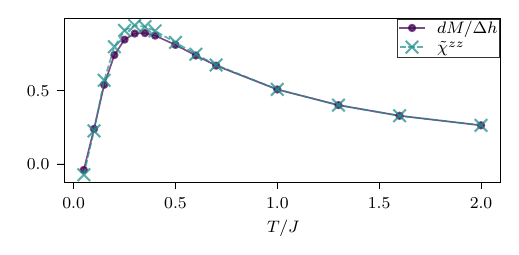}
	\caption{Confirmation of the relation \eqref{eq:chitilde} between field derivatives of the magnetization and the susceptibility computed with $U(1)$-pm-fRG for the dimer. The model parameters are $J^z=1,\ J^{\perp}=0.5$, $h=1$ and we set $\Delta h = 0.002$. Lines are guides to the eye.}
	\label{fig_chi_dM}
\end{figure}

Following the practice in the development of pm-fRG \cite{Niggemann2021,schneider2024tamingspins,Schneider2024_Tflow} we begin with the simple model of two spins $j=1,2$ interacting with Hamiltonian \eqref{eq_XXZ--Z}. Though analytically solvable, from the viewpoint of pm-fRG this dimer model possesses already the full complexity which makes it an excellent system for benchmarking. In Fig.~\ref{AFM_Dimer} we show the magnetization and static susceptibilities for the AFM Heisenberg and Ising case in various $Z$-fields, see panels (a) and (b), respectively. As in standard pm-fRG\cite{Niggemann2021} the $U(1)$-pm-fRG results are excellent at large temperatures $T \gtrsim J$ and start to deviate strongly from the exact solution at model-dependent temperature scales usually around $T \simeq J/3$. In both cases, we also verified the anticipated scaling of the error at large $T/J$. We observe that the fRG performs better for the Ising than for the Heisenberg model, in the latter deviations are especially severe at low temperatures if the ground state is degenerate (for $h = 1$). As we will show later, better accuracy of the fRG can be expected in higher spatial dimensions by full incorporation of the mean-field equations\cite{Mueller2023}. Finally, we perform the following consistency check\cite{Noculak2024} between the static susceptibilities calculated from Eq.~\eqref{eq_suceptibility_zz} using the fermionic four-point vertex and appropriate field-derivatives of local magnetizations. We break the dimer's inversion symmetry by small $\Delta h_1 = -\Delta h_2\equiv \Delta h$ and obtain from $\Delta M \equiv M_1-M_2$ the following relation valid in the linear response regime $\Delta h \rightarrow 0$,
\begin{equation} 
\frac{\Delta M}{\Delta h} = \frac{M_1-M_2}{\Delta h} = \chi^{zz}_{11} - \chi^{zz}_{12} - \chi^{zz}_{21} + \chi^{zz}_{22} \equiv \tilde{\chi}^{zz}.\label{eq:chitilde}
\end{equation}
In Fig.~\ref{fig_chi_dM} we confirm that this relation is indeed fulfilled to few percent accuracy for a dimer with $J^z=1$, $J^\perp=0.5$, $h=1$ and $\Delta h=0.002$.


%%%%%%%%%%%%%%%%%%%%%%%%%%%%%%%%%%%%%%%%%%%%%
\subsection{FM Heisenberg model on square lattice}
\label{FM Heisenberg Square Lattice}
\begin{figure}
	\centering
	\includegraphics[scale=1.0]{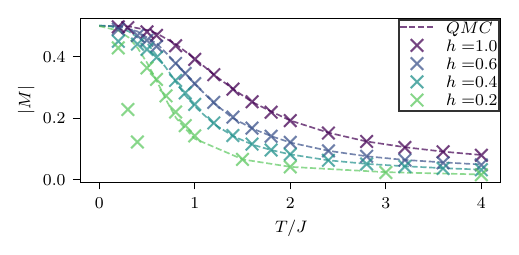}
	\caption{Magnetization $|M|$ for the FM Heisenberg model on the square lattice for various applied magnetic fields $h$. Cross markers show fRG results and lines denote QMC results\cite{Henelius_2000}.}
	\label{fig_square_lat_mag}
\end{figure}

We now turn to extended systems and demonstrate the $U(1)$-pm-fRG's ability to capture magnetization curves $M(T)$. We compare our results for the FM Heisenberg model on the square lattice to error controlled QMC results of Ref.~\onlinecite{Henelius_2000}. The fRG result is displayed in Fig.~\ref{fig_square_lat_mag} and shows good agreement with QMC until low temperatures where the truncation of higher order vertices is not justified any longer. We observe that the quality of the fRG results improves with increasing magnetic field $h$. This is to be expected since a larger $h$ diminishes the relative weight of the interacting term in the Hamiltonian $H$. Following the line for $h=0.2$ to low temperatures, a drastic and unphysical dip in $M(T)$ is observed. This is traced back to sharp changes in some vertex components and an associated numerical instability in the fRG flow. 
Such a behavior often arises when the slope of $M(T)$ is large. This generally occurs at low $h$ and large $L$.

%%%%%%%%%%%%%%%%%%%%%%%%%%%%%%%%%%%%%%%%%%%%
\subsection{Spontaneous magnetization on cubic lattice Heisenberg model}
\label{Spontaneous_Magentization}
In three spatial dimensions (short-range) XXZ models can feature long-range order and spontaneous breaking of spin-rotation and time-reversal symmetries below a critical temperature $T_c$. Previous applications of the pm-fRG\cite{Niggemann2022,sbierski2023,Schneider2024_Tflow} have shown the method's potential to determine $T_c$ within a few percent accuracy from exact results (when the latter are available). One of the new features of the $U(1)$-pm-fRG is the ability to enter the spontaneously ordered phase [if it preserves $U(1)$ spin rotation symmetry] and compute the magnetization $M(T)$ across the phase transition. However, a small but finite symmetry breaking field $h$ is required. 

\begin{figure}
	\centering
	\includegraphics[scale=1.0]{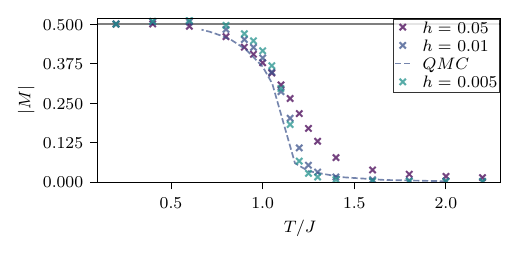}
	\caption{Magnetization $|M|$ for the XXZ Heisenberg model ($J^z=-1, J^{\perp}=0.3$) on the cubic lattice with various small seed fields $h$. Cross markers show fRG results with $L=10$ as the radius of the correlation ball. A dashed line shows error controlled QMC results for $h=0.01$ and a cube of linear extent of ten sites and periodic boundary conditions. Finite-size corrections are negligible on the scale of the plot. The small overshoot $M \gtrsim 0.5$ for small temperatures is a numerical artifact related to truncation of frequency sums.}
	\label{fig_spontaneous_mag}
\end{figure}

To test these capabilities we turn to the mixed FM-AFM cubic lattice XXZ model with interactions $J^z=-1, J^{\perp}=0.3$ and plot the $Z$-magnetization $M(T)$ with various small fields $h$. The results in Fig.~\ref{fig_spontaneous_mag} show the expected sharp increase in magnetization around $T_c$. The onset of magnetization sharpens with decreasing $h$. For $h=0.01$ we also compare to QMC results obtained with the worm-type QMC code of Ref.~\onlinecite{Pollet_code1,Pollet_code2} and observe good agreement. 

%%%%%%%%%%%%%%%%%%%%%%%%%%%%%%%%%%%%%%%%%%%%%
%%%%%%%%%%%%%%%%%%%%%%%%%%%%%%%%%%%%%%%%%%%%%
\section{Application: Triangular lattice}
\label{sec:triangular}
We finally apply the $U(1)$-pm-fRG developed above to real materials studied in recent experiments. In particular, we consider $S=1/2$ compounds with triangular lattice structure and frustrated in-plane AFM couplings $J^\perp>0$ not amenable to QMC simulations due to the sign problem\cite{Sandvik2010}.    

%%%%%%%%%%%%%%%%%%%%%%%%%%%%%%%%%%%%%%%%%%%%%
\subsection{Magnetization in CeMgAl$_{11}$O$_{19}$}
\label{Magnetization in CeMgAl$_{11}$O$_{19}$}
Our first application considers the recently investigated rare-earth hexaaluminate material $\mathrm{CeMgAl}_{11}\mathrm{O}_{19}$. Using inelastic neutron spectra in fully polarizing magnetic out-of-plane fields $h$, Gao et al.\cite{gao2024} determined the Hamiltonian to be well approximated as nearest neighbor XXZ with $J^{\perp} = 0.6469K$ and $J^{z} = -0.2784K$. The large anisotropy is due to spin-orbit coupling. Crucially, the ratio of both couplings is close to the point $J^\perp/J^z=-0.5$ which separates magnetically ordered FM and coplanar 120° phases. At this point, the model features and exactly solvable and highly degenerate ``spin-liquid" ground state \cite{Momoi192}. In Ref.~\onlinecite{gao2024}, this insight was used to model the measured $h=0$ inelastic neutron scattering spectrum of $\mathrm{CeMgAl}_{11}\mathrm{O}_{19}$.

\begin{figure}
	\centering
	\includegraphics[scale=1.0]{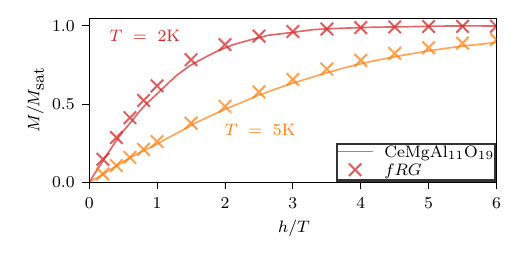}
	\caption{Consistency check of the proposed XXZ--Z model for $\mathrm{CeMgAl}_{11}\mathrm{O}_{19}$ using the magnetization data $M(h)$ in units of the saturated magnetization (lines) from Ref.~\onlinecite{gao2024}: The crosses show results from the $U(1)$-pm-fRG applied to the triangular lattice XXZ--Z model with $J^{\perp} = 0.6469K$, $J^{z} = -0.2784K$ and $g^z=3.66$ at temperatures $T=2$K and $5$K.}
	\label{fig_M_Gao}
\end{figure}

Here, by employing the $U(1)$-pm-fRG we verify that the proposed model indeed accurately reproduces the experimental out-of-plane magnetization data provided in Ref.~\onlinecite{gao2024} for $T=2K$ and $5K$, see Fig.~\ref{fig_M_Gao}. We use the Landé factor $g^z=3.66$ reported from an ESR measurement\cite{gao2024} to model the Zeeman term as $H_Z=g^z\mu_{B}\sum_j h S^z_j$ where $\mu_B$ is the Bohr magneton.  

%%%%%%%%%%%%%%%%%%%%%%%%%%%%%%%%%%%%%%%%%%%%%
\subsection{Three-sublattice spin solid in Na$_2$BaCo(PO$_4$)$_2$}
\label{Critical Temperature of Na$_2$BaCo(PO$_4$)$_2$ In An External Magnetic Field}

As a second application of the $U(1)$-pm-fRG we consider the material Na$_2$BaCo(PO$_4$)$_2$ recently studied in Refs.~\onlinecite{Zhong2019,Li2020,gao2024b,Gao2022,Sheng2024} and identified to host a spin supersolid phase\cite{Xiang2024}. The Co$^{2+}$ ions are host to effective $S=1/2$ spins. The nearest neighbor XXZ--Z model parameters were estimated as\cite{Xiang2024} $J^{\perp} = 0.8K$, $J^{z} = 1.48K$ with $g^z=4.89$ the Landé factor. 

\begin{figure}
	\centering
	\includegraphics[scale=1.0]{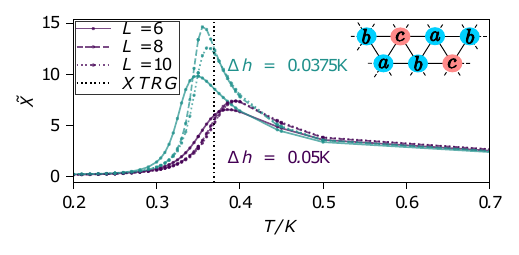}
	\caption{Susceptibility $\tilde{\chi}^{zz}$ from Eq.~\eqref{eq_chi3subl} detects the transition to three-sublattice order (inset) in the XXZ--Z triangular lattice model for Na$_2$BaCo(PO$_4$)$_2$ with $J^{\perp} = 0.8K$, $J^{z} = 1.48K$, $h=2.465K$ studied in Ref.~\onlinecite{Xiang2024}. The two values used for the small symmetry-breaking field $\Delta h$ are given in the panel. Even smaller symmetry breaking field would cause numerical instabilities in the fRG flow. The vertical line shows the XTRG results from Ref.~\onlinecite{Xiang2024}.}
	\label{fig_results_chi}
\end{figure}

A spin supersolid is characterized by the simultaneous breaking of lattice translation symmetry and $U(1)$-spin rotation symmetry. However, in the model at hand this only occurs at very small temperatures $<0.2K$ and thus remains out of reach for the fRG. Instead we consider the transition to three-sublattice up-up-down spin solid order also found experimentally in Ref.~\onlinecite{Xiang2024}, see the inset of Fig.~\ref{fig_results_chi} for a sketch. We set $h=2.465K$ (corresponding to $0.75T$) and consider the susceptibility relating the order parameter $(M_a+M_b)/2-M_c$ to an additional symmetry-breaking seed field $\Delta h_a = \Delta h_b = -\Delta h_c \equiv \Delta h$ (which we keep finite for numerical stability across $T_c$),
\begin{equation} 
\begin{split}
\tilde{\chi}^{zz} &= \sum_j \left[ \left( \chi^{zz}_{1_aj_a} + \chi^{zz}_{1_aj_b} -\chi^{zz}_{1_aj_c} \right)/2 \right. \\
& + \left( \chi^{zz}_{1_bj_a} + \chi^{zz}_{1_bj_b} -\chi^{zz}_{1_bj_c} \right)/2\\
& - \left( \chi^{zz}_{1_cj_a} + \chi^{zz}_{1_cj_b} -\chi^{zz}_{1_cj_c} \right) \left. \right].\label{eq_chi3subl}
\end{split}
\end{equation}
Here $j_s$ labels sub-lattice sites $s=a,b,c$ for lattice site $j$ and the $\chi^{zz}$ on the right-hand side are computed using Eq.~\eqref{eq_suceptibility_zz}. In Fig.~\ref{fig_results_chi} we show the resulting peak structure of $\tilde{\chi}^{zz}(T)$ which sharpens with increased correlation distance $L=6,8,10$. The center of the peak indicates the transition temperature $T_c$, does not vary much with $\Delta h$ and is in good agreement with experimental and exponential tensor renormalization group (XTRG) results\cite{ChenExponentialThermalTensor2018,Xiang2024} (vertical line). The main advantage of our $U(1)$-pm-fRG over the XTRG simulation of Ref.~\onlinecite{Xiang2024} is the ensured convergence in system size. While we use a correlation disc of radius $L\leq10$ embedded in an infinite lattice, the XTRG\cite{Xiang2024} operates on cylinders of relatively small width $W=6$.

%%%%%%%%%%%%%%%%%%%%%%%%%%%%%%%%%%%%%%%%%%%%%%%%%%%%%%%%%%%%%%%%
%%%%%%%%%%%%%%%%%%%%%%%%%%%%%%%%%%%%%%%%%%%%%%%%%%%%%%%%%%%%%%%%
\section{Summary and Outlook}
\label{summary}
We have introduced a $U(1)$-pm-fRG scheme based on a symmetry adapted mixed fermionic spin representation for XXZ--Z type spin $S=1/2$ models. 
%This representation is based on the faithful $SO(3)$ pseudo-Majorana representation\fb{, which could be generalized to higher spins $S>1/2$, but only $S=1/2$ and $S = 3/2$ are faithful representations\cite{hagymasi2024,Schaden2023s1}}. 
The advantage of the mixed representation is that it diagonalizes the non-interacting Hamiltonian (and Green's function). This makes fRG calculations with an external magnetic field feasible and allows for the calculation of magnetization and access to the ordered phase. We have derived relevant observables and the fRG flow equations in the usual Katanin truncation with a smooth Matsubara frequency cutoff. The implementation of the more efficient temperature flow scheme recently introduced in the context of pm-fRG\cite{Schneider2024_Tflow} is left for future work. Here, technical complications are expected since the presence of magnetic fields require large frequency boxes, which could be made feasible by applying vertex compression schemes \cite{rohshap2024}. 

We also provided compelling benchmark checks against established exact and QMC results. For applications to real materials, we chose the highly frustrated triangular lattice in magnetic fields. As a first example, in the case of CeMgAl$_{11}$O$_{19}$ we show excellent agreement of fRG simulations of magnetization curves of the proposed model against experimental results. Second, we reproduced the experimentally determined critical temperature for the transition into the up-up-down three-sublattice spin solid phase in Na$_2$BaCo(PO$_4$)$_2$.

We expect that the $U(1)$-pm-fRG will be especially useful for the application to three-dimensional frustrated XXZ--Z magnets where alternative methods are scarce. In addition, it is also straightforward to adapt our approach to other spin models with $U(1)$-symmetry. This includes, for example, the Jaynes-Cummings model of quantum optics\cite{larsonJaynesCummings2024} after the bosonic degrees of freedom have been traced out to produce a retarded spin-spin interaction.
It would also be interesting to generalize the $U(1)$-pm-fRG for XXZ-Z models to larger spins, especially to $S=3/2$ where a faithful pseudo-Majorana representation exists\cite{Schaden2023s1}. The pm-fRG was also recently applied to a $S=1$ model (without magnetic fields) where however the influence of the inevitable unphysical states had to be mitigated\cite{hagymasi2024}.

A numerical implementation of the $U(1)$-pm-fRG is available at \url{https://gitlab.com/Frederic.Bippus/U1-pm-fRG.git}

%%%%%%%%%%%%%%%%%%%%%%%%%%%%%%%%%%%%%%%%%%%%%%%%%%%%%%%%%%
\begin{acknowledgments}
We thank Jan von Delft, Johannes Reuther and the authors of Ref.~\onlinecite{gao2024} for discussions. F.~B.~acknowledges support by the SFB Q-M\&S (FWF project ID F86). B.~Sch.~and B.~Sb.~acknowledge support from DFG grant no.~524270816. B.~Sb.~acknowledges support from DFG through the Research Unit FOR 5413/1 (grant no.~465199066).  B.~Sch.~acknowledges funding from the Munich Quantum Valley, supported by the
Bavarian state government with funds from the Hightech Agenda Bayern Plus. The authors gratefully acknowledge the Gauss Centre for Supercomputing e.V. (www.gauss-centre.eu) for funding this project by providing computing time through the John von Neumann Institute for Computing (NIC) on the GCS Supercomputer JUWELS at Jülich Supercomputing Centre (JSC).
%For the purpose of open access, the authors have applied a CC BY-NC-SA public copyright license to any Author Accepted Manuscript version arising from this submission.
\end{acknowledgments}

%%%%%%%%%%%%%%%%%%%%%%%%%%%%%%%%%%%%%%%%%%%%%%%%%%%%%%%%
%%%%%%%%%%%%%%%%%%%%%%%%%%%%%%%%%%%%%%%%%%%%%%%%%%%%%%%%
\appendix

%%%%%%%%%%%%%%%%%%%%%%%%%%%%%%%%%%%%%%%%%%%%%%%%%%%%%%%%
%%%%%%%%%%%%%%%%%%%%%%%%%%%%%%%%%%%%%%%%%%%%%%%%%%%%%%%%
\section{Details on Green's functions}\label{Appendix_G0}

We explain how the expressions for the full Green's function Eq.~\eqref{eq_G_matrix} are obtained from the action in Eq.~\eqref{eq_action_XXZ}. In the convention of Ref.~\onlinecite{Kopitz2010} the matrix components of the non-interacting Green's function in the superfield basis read
\begin{subequations} 
\begin{align}
	\left[\mathbf{G}_0^{-1}\right]_{\psi_1\bar{\psi}_2} &=  \beta \delta_{j_1j_2}\delta_{\omega_1,-\omega_2} \left( i\omega_1 - h_{j_1} \right) \nonumber \\
    &\equiv \beta \delta_{j_1j_2}\delta_{\omega_1,-\omega_2} G^{-1}_{0,\psi_1\bar{\psi}_2}(\omega_1), \\
	\left[\mathbf{G}_0^{-1}\right]_{\bar{\psi}_1\psi_2} &= \beta \delta_{j_1j_2}\delta_{\omega_1,-\omega_2} \left( i\omega_1 + h_{j_1}  \right) \nonumber \\
    &\equiv \beta \delta_{j_1j_2}\delta_{\omega_1,-\omega_2}G^{-1}_{0,\bar{\psi}_1\psi_2}(\omega_1), \\
	\left[\mathbf{G}_0^{-1}\right]_{\zeta_1\zeta_2} &=  \beta \delta_{j_1j_2}\delta_{\omega_1,-\omega_2} \left( i\omega_1\right) \nonumber \\
    &\equiv \beta \delta_{j_1j_2}\delta_{\omega_1,-\omega_2} G^{-1}_{0,\zeta_1\zeta_2}(\omega_1).
\end{align}
\end{subequations} 
The inversion of the above equations requires special care in the frequency- and particle sectors,
\begin{subequations} 
\begin{align}\label{eq_g0}
	\left[\mathbf{G}_0 \right]_{\psi_1\bar{\psi}_2} &=  - \frac{1}{\beta} \delta_{j_1j_2}\delta_{\omega_1,-\omega_2} \frac{1}{i\omega_1 - h_{j_1} } \nonumber \\
    &\equiv - \frac{1}{\beta} \delta_{j_1j_2}\delta_{\omega_1,-\omega_2} G^{0}_{\psi_1\bar{\psi}_2}(\omega_1), \\
	\left[\mathbf{G}_0\right]_{\bar{\psi}_1\psi_2} &= - \frac{1}{\beta} \delta_{j_1j_2}\delta_{\omega_1,-\omega_2} \frac{1}{ i\omega_1 + h_{j_1} } \nonumber \\
    &\equiv - \frac{1}{\beta} \delta_{j_1j_2}\delta_{\omega_1,-\omega_2}G^0_{\bar{\psi}_1\psi_2}(\omega_1), \\
	\left[\mathbf{G}_0\right]_{\zeta_1\zeta_2} &=  - \frac{1}{\beta} \delta_{j_1j_2}\delta_{\omega_1,-\omega_2} \frac{1}{i\omega_1} \nonumber \\
    &\equiv - \frac{1}{\beta} \delta_{j_1j_2}\delta_{\omega_1,-\omega_2} G^0_{\zeta_1\zeta_2}(\omega_1).
\end{align}
\end{subequations} 
The propagator matrices are anti-symmetric,
\begin{equation}
	G^{0}_{12}(\omega_1) = - G^{0}_{21}(\omega_2) = - G^{0}_{21}(-\omega_1).
\end{equation}
The full propagator is given by Dyson's equation,
\begin{equation}
		\left[\mathbf{G}\right]_{12} = \left[\mathbf{G}^0\right]_{12} + \left[\mathbf{G}^0\right]_{13}\left[\mathbf{\Sigma}\right]_{34}\left[\mathbf{G}\right]_{42},
\end{equation}
from which we obtain Eq.~\eqref{eq_G_matrix}.

%%%%%%%%%%%%%%%%%%%%%%%%%%%%%%%%%%%%%%%%%%%%%%%%%%%%%%%%
%%%%%%%%%%%%%%%%%%%%%%%%%%%%%%%%%%%%%%%%%%%%%%%%%%%%%%%%
\section{Detailed flow equations}\label{Appendix_Flow_Equations}

%%%%%%%%%%%%%%%%%%%%%%%%%%%%%%%%%%%%%%%%%%%%%%%%%%%%%%%%
\subsection{Four-point vertex}\label{4flowequations}

\begin{figure*}
	\centering
	\includegraphics[scale=1]{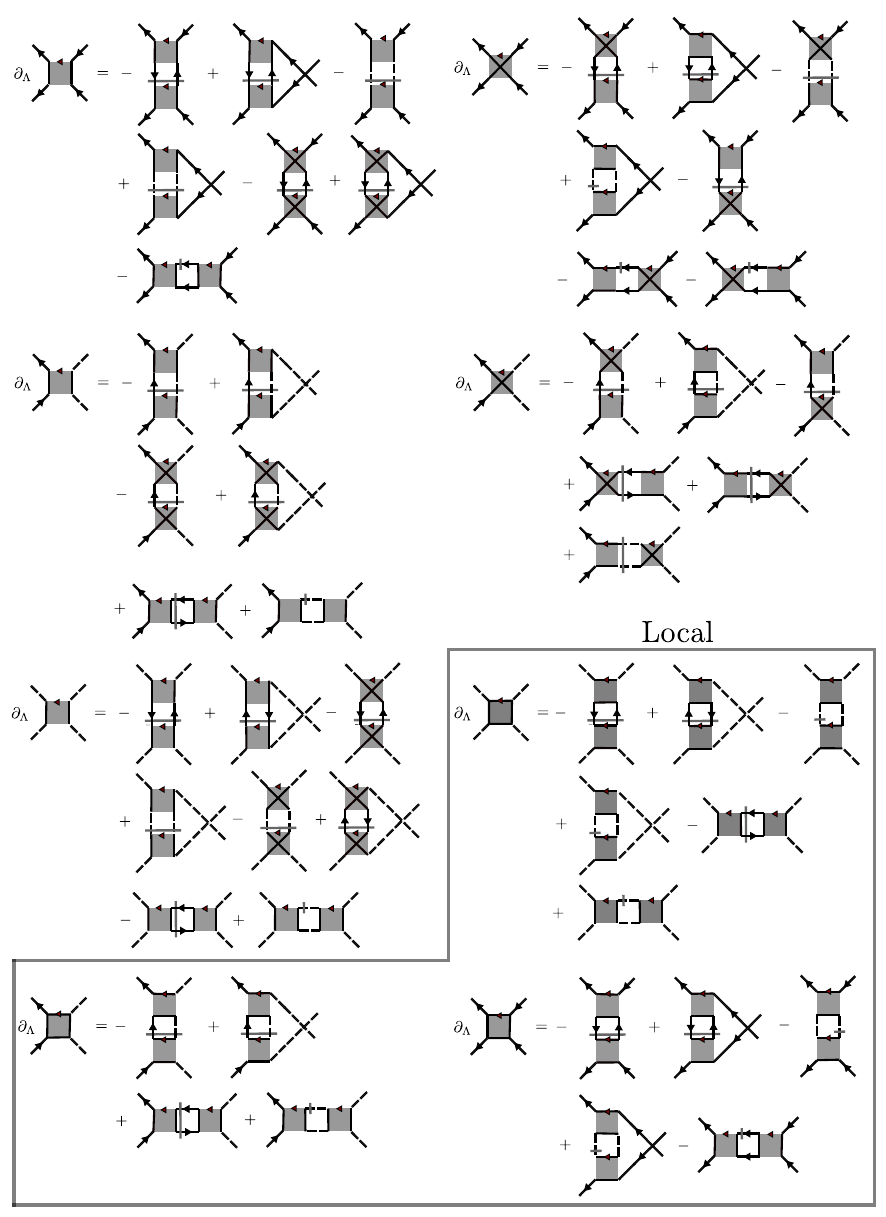}
	\caption{ Diagrammatic representation of the four point flow equations for $\del_{\Lambda}\Gamma_{\psi\psi,ij},\ \del_{\Lambda}\Gamma^{\times}_{\psi\psi,ij},\ \del_{\Lambda}\Gamma_{\psi\zeta,ij},\ \del_{\Lambda}\Gamma^{\times}_{\psi\zeta,ij},\ \del_{\Lambda}\Gamma_{\zeta\zeta,ij}$ and the local terms $\del_{\Lambda}\Gamma_{\psi\psi,ii},\ \del_{\Lambda} \Gamma_{\psi\zeta,ii},\ \del_{\Lambda} \Gamma_{\zeta\zeta,ii}$. Here we skip frequency and site labels as these are given in Eqns.~\eqref{eq_pi_terms}-\eqref{eq_local_4_flow}.
 }
	\label{fig_feynmandiagrams}
\end{figure*}

We present the flow equations of the one-line irreducible four-point vertices. The six-point vertex contributions are truncated due to numerical limitations. A method to partially include them without a significant increase in numerical cost is introduced in Appendix~\ref{Katanin}. In the following, we drop all $\Lambda$ labels for brevity and begin by defining internal propagator pairs,
\begin{subequations} 
\begin{align}\label{eq_pi_terms}
	\Pi_{\psi\psi,ij}\left(\omega,\omega+s\right) & \equiv \frac{1}{2\beta} \left( \dot{G}_{\psi,i}\left(\omega \right) G_{\psi,j}\left(\omega+s \right) \right. \nonumber \\ 
    &+ \left. G_{\psi,i}\left(\omega \right) \dot{G}_{\psi,j}\left(\omega+s \right)  \right),\\
	\Pi_{\psi\zeta,ij}\left(\omega,\omega+s\right) & \equiv \frac{1}{2\beta} \left( \dot{G}_{\psi,i}\left(\omega \right) G_{\zeta,j}\left(\omega+s \right) \right. \nonumber \\ 
    &+ \left. G_{\psi,i}\left(\omega \right) \dot{G}_{\zeta,j}\left(\omega+s \right)  \right),\\
	\Pi_{\zeta\zeta,ij}\left(\omega,\omega+s\right) & \equiv \frac{1}{2\beta} \left( \dot{G}_{\zeta,i}\left(\omega \right) G_{\zeta,j}\left(\omega+s \right) \right. \nonumber \\ 
    &+ \left. G_{\zeta,i}\left(\omega \right) \dot{G}_{\zeta,j}\left(\omega+s \right) \right).
\end{align}
\end{subequations} 
We group the contributions to the right-hand sides of the four-point vertex flow equations by their properties as follows. Terms with an internal site sum are called $X$-terms. For these terms, the flow equations for the local and bi-local case are equal and need not be distinguished below. In contrast, terms without internal sums are called $\tilde{X}$-terms, for these terms the local case indicated by site-indices $ii$ needs to be distinguished from the non-local case indicated by site indices $ij$. For non crossed $\tilde{X}$ terms $t-u$ symmetries exist, their application to Eqns.~\eqref{eq_1}-\eqref{eq_2} significantly reduces the numerical cost. For the crossed components ($\times$), this symmetry does not exist. Here we start by stating the flow equations analytically in terms of these components before stating the components explicitly. A diagrammatic representation of the flow equations is displayed in Fig.~\ref{fig_feynmandiagrams}.

\begin{widetext}
The non-local flow equations (for $i\neq j$) are
\begin{subequations} 
\begin{align}
	\del_{\Lambda} \Gamma_{\psi\psi,ij}\left(s,t,u\right) & = X_{\psi\psi,ij}\left(s,t,u\right) + \tilde{X}_{\psi\psi,ij}\left(s,t,u\right) - \tilde{X}_{\psi\psi,ij}\left(s,u,t\right),\\ \label{eq_1}
	\del_{\Lambda} \Gamma_{\zeta\zeta,ij}\left(s,t,u\right) & = X_{\zeta\zeta,ij}\left(s,t,u\right) + \tilde{X}_{\zeta\zeta,ij}\left(s,t,u\right) - \tilde{X}_{\zeta\zeta,ij}\left(s,u,t\right),\\
	\del_{\Lambda} \Gamma_{\psi\zeta,ij}\left(s,t,u\right) & = X_{\psi\zeta,ij}\left(s,t,u\right) + \tilde{X}_{\psi\zeta,ij}\left(s,t,u\right) - \tilde{X}_{\psi\zeta,ij}\left(s,u,t\right),\\ \label{eq_2}
	\del_{\Lambda} \Gamma^{\times}_{\psi\psi,ij}\left(s,t,u\right) & = X^{\times}_{\psi\psi,ij}\left(s,t,u\right) + \tilde{X}^{\times}_{\psi\psi,ij}\left(s,t,u\right),\\
	\del_{\Lambda} \Gamma^{\times}_{\psi\zeta,ij}\left(s,t,u\right) & = X^{\times}_{\psi\zeta,ij}\left(s,t,u\right) + \tilde{X}^{\times}_{\psi\zeta,ij}\left(s,t,u\right),
\end{align}
and the local flow equations are given by
\begin{align}
	\del_{\Lambda} \Gamma_{\psi\psi,ii}\left(s,t,u\right) & = \del_{\Lambda} \Gamma^{\times}_{\psi\psi,ii}\left(s,t,u\right) = X_{\psi\psi,ii}\left(s,t,u\right) + \tilde{X}_{\psi\psi,ii}\left(s,t,u\right) - \tilde{X}_{\psi\psi,ii}\left(s,u,t\right), \\
	\del_{\Lambda} \Gamma_{\zeta\zeta,ii}\left(s,t,u\right) & = X_{\zeta\zeta,ii}\left(s,t,u\right) + \tilde{X}_{\zeta\zeta,ii}\left(s,t,u\right) - \tilde{X}_{\zeta\zeta,ii}\left(s,u,t\right),\\
	\del_{\Lambda} \Gamma_{\psi\zeta,ii}\left(s,t,u\right) & = \del_{\Lambda} \Gamma^{\times}_{\psi\zeta,ii}\left(s,t,u\right) = X_{\psi\zeta,ii}\left(s,t,u\right) + \tilde{X}_{\psi\zeta,ii}\left(s,t,u\right) - \tilde{X}_{\psi\zeta,ii}\left(s,u,t\right). \label{eq_local_4_flow}
\end{align}
\end{subequations} 
The $X$-contributions to the flow equations read,
\begin{subequations} 
\begin{equation}\label{eq_Xccij}
	\begin{split}
		X_{\psi\psi,ii(j)}&\left(s,t,u\right)
		=\\
		& - \sum_{\omega} \sum_{k} \Pi_{\psi\psi,kk}\left(-\omega,\omega+s \right)
		\Gamma_{\psi\psi,ki(j)}\left(s, -\omega+\omega_{3}, -\omega+\omega_{4} \right) \Gamma_{\psi\psi,ik}\left(s, -\omega-\omega_{2}, \omega+\omega_{1} \right), \\				
	\end{split}
\end{equation}
\begin{equation}\label{eq_Xmmij}
	\begin{split}
		X_{\zeta\zeta,ii(j)}&\left(s,t,u\right)
		=\\
		&\quad 2 \sum_{\omega} \sum_{k} \Pi_{\psi\psi,kk}\left(\omega,\omega+s\right) \Gamma_{\psi\zeta,ki(j)}\left(s, \omega-\omega_{4}, \omega-\omega_{3} \right) \Gamma_{\psi\zeta,ki}\left(-s, \omega+\omega_{1}, \omega+\omega_{2} \right) \\
		&+ 
		\sum_{\omega} \sum_{k} \Pi_{\zeta\zeta,kk}\left(\omega,\omega+s\right) \Gamma_{\zeta\zeta,ki(j)}\left(s, -\omega+\omega_{3}, -\omega+\omega_{4} \right) \Gamma_{\zeta\zeta,ki}\left(-s, -\omega-\omega_{2}, -\omega-\omega_{1} \right), \\						
	\end{split}
\end{equation}
\begin{equation}\label{eq_Xcmij}
	\begin{split}
		X_{\psi\zeta,ii(j)}&\left(s,t,u\right)
		=\\
		& - 2 \sum_{\omega} \sum_{k} \Pi_{\psi\psi,kk}\left(\omega,\omega+s\right) \Gamma_{\psi\zeta,ki(j)}\left(s, \omega-\omega_{4}, \omega-\omega_{3} \right) \Gamma^{\times}_{\psi\psi,ik}\left(\omega+\omega_{1}, s, -\omega-\omega_{2} \right) \\						
		&+
		\sum_{\omega}  \sum_{k} \Pi_{\zeta\zeta,kk}\left(\omega,\omega+s\right)  \Gamma_{\zeta\zeta,ki(j)}\left(s, -\omega+\omega_{3}, -\omega+\omega_{4} \right) \Gamma_{\psi\zeta,ik}\left(s, -\omega-\omega_{2}, \omega+\omega_{1} \right) .\\ 																		
	\end{split}
\end{equation}
The $X^{\times}$-contributions are
\begin{equation}\label{eq_Xxccij}
	\begin{split}
		X^{\times}_{\psi\psi,ij}&\left(s,t,u\right)
		=\\
		&- 2 \sum_{\omega}\sum_{k} \Pi_{\psi\psi,kk}\left(\omega,\omega-t\right) \Gamma^{\times}_{\psi\psi,ik}\left( \omega-\omega_3,t,-\omega + \omega_1\right) \Gamma^{\times}_{\psi\psi,jk}\left(\omega+\omega_2,-t, -\omega-\omega_4\right) \\
		& -
		\sum_{\omega}\sum_{k}  \Pi_{\zeta\zeta,kk}\left(\omega,\omega-t\right) \Gamma_{\psi\zeta,ik}\left(t, \omega-\omega_3, -\omega+\omega_1\right) \Gamma_{\psi\zeta,jk}\left(-t, \omega+\omega_2, -\omega-\omega_4\right), \\
	\end{split}
\end{equation}
\begin{equation}\label{eq_Xxcmij}
	\begin{split}
		X^{\times}_{\psi\zeta,ij}&\left(s,t,u\right)
		=\\
		& 2 \sum_{\omega}\sum_{k} \Pi_{\psi\zeta,kk}\left(\omega,\omega-t\right)  \Gamma^{\times}_{\psi\zeta,ik}\left(-\omega+\omega_1,t, \omega-\omega_3\right) \Gamma^{\times}_{\psi\zeta,kj}\left( \omega+\omega_2,t, \omega+\omega_4\right).\\		
	\end{split}
\end{equation}
\end{subequations} 
For the above definitions, local and non-local cases have the same expression. For terms without internal lattice site sums $\tilde{X}$, this is not the case. Their non-local contributions read
\begin{subequations} 
\begin{equation}\label{eq_Xtccij}
	\begin{split}
		\tilde{X}_{\psi\psi,ij}&\left(s,t,u\right)
		=  \\			
		& \sum_{\omega}\left\{ 2 \Pi_{\psi\psi,ij}\left(\omega,\omega-t \right) \Gamma^{\times}_{\psi\psi,ij}\left( \omega-\omega_3,-\omega + \omega_1,t\right) \Gamma_{\psi\psi,ij}\left(\omega+\omega_2,-t, -\omega-\omega_4\right) \right. \\			
		& + 
		2 \Pi_{\psi\psi,ji}\left(\omega,\omega-t \right) \Gamma_{\psi\psi,ij}\left( \omega-\omega_3,t,-\omega + \omega_1\right) \Gamma^{\times}_{\psi\psi,ij}\left(\omega+\omega_2, -\omega-\omega_4,-t\right) \\					
		&+ \left.
		2 \Pi_{\zeta\zeta,ij}\left(\omega,\omega-t \right) \Gamma^{\times}_{\psi\zeta,ij}\left(t, -\omega+\omega_1,\omega-\omega_3\right) \Gamma^{\times}_{\psi\zeta,ij}\left(-t, \omega+\omega_2, -\omega-\omega_4\right)\right\},\\
	\end{split}
\end{equation}
\begin{equation}\label{eq_Xtxccij}
	\begin{split}
		\tilde{X}^{\times}_{\psi\psi,ij}&\left(s,t,u\right)
		=  \\			
		& \sum_{\omega} \left\{ \Pi_{\psi\psi,ij}\left(-\omega,\omega+s\right) \Gamma^{\times}_{\psi\psi,ij}\left(s, -\omega+\omega_{3}, -\omega+\omega_{4} \right) \Gamma^{\times}_{\psi\psi,ij}\left(s, \omega+\omega_{1},-\omega-\omega_{2} \right) \right. \\
		&+  
		\Pi_{\psi\psi,ji}\left(-\omega,\omega+s\right) \Gamma^{\times}_{\psi\psi,ji}\left(s, -\omega+\omega_{4}, -\omega+\omega_{3} \right) \Gamma^{\times}_{\psi\psi,ij}\left(s, -\omega-\omega_{2}, \omega+\omega_{1} \right) \\									
		& + 
		2 \Pi_{\psi\psi,ij}\left(\omega,\omega+u\right) \Gamma_{\psi\psi,ji}\left( \omega-\omega_3,-\omega + \omega_2,-u\right) \Gamma_{\psi\psi,ij}\left(\omega+\omega_1, -\omega-\omega_4,u\right) \\			
		& + 
		2 \Pi_{\psi\psi,ji}\left(\omega,\omega+u\right)  \Gamma^{\times}_{\psi\psi,ji}\left( \omega-\omega_3,-\omega + \omega_2,-u\right) \Gamma^{\times}_{\psi\psi,ij}\left(\omega+\omega_1, -\omega-\omega_4,u\right) \\			
		&+ \left.
		2 \Pi_{\zeta\zeta,ij}\left(\omega,\omega+u\right) \Gamma^{\times}_{\psi\zeta,ji}\left(-u, \omega-\omega_3, -\omega+\omega_2\right) \Gamma^{\times}_{\psi\zeta,ij}\left(u, \omega+\omega_1, -\omega-\omega_4\right) \right\},\\		
	\end{split}
\end{equation}
\begin{equation}\label{eq_Xtmmij}
	\begin{split}
		\tilde{X}_{\zeta\zeta,ij}&\left(s,t,u\right)
		=  \\			
		& \sum_{\omega}\left\{ 
		2 \Pi_{\psi\psi,ij}\left(\omega,\omega-t\right) \Gamma^{\times}_{\psi\zeta,ji}\left(-t, \omega-\omega_1, \omega-\omega_3\right) \Gamma^{\times}_{\psi\zeta,ij}\left(t, \omega+\omega_2, \omega+\omega_4\right)\right.\\			
		& + 
		2 \Pi_{\psi\psi,ji}\left(\omega,\omega-t\right) \Gamma^{\times}_{\psi\zeta,ij}\left(-t, \omega-\omega_3, \omega-\omega_1\right) \Gamma^{\times}_{\psi\zeta,ji}\left(t, \omega+\omega_4, \omega+\omega_2\right)\\									
		& - 
		2 \Pi_{\zeta\zeta,ij}\left(\omega,\omega-t\right) \left. \Gamma_{\zeta\zeta,ij}\left(  -\omega+\omega_1,-\omega+\omega_3,-t\right)  \Gamma_{\zeta\zeta,ji}\left(-\omega-\omega_2,t, -\omega-\omega_4\right)  \right\},\\
	\end{split}
\end{equation}
\begin{equation}\label{eq_Xtcmij}
	\begin{split}
		\tilde{X}_{\psi\zeta,ij}&\left(s,t,u\right)
		=  \\			
		&\sum_{\omega} \left\{ - 
		2 \Pi_{\psi\zeta,ij}\left(\omega,\omega-t\right) \Gamma_{\psi\zeta,ij}\left(-\omega+\omega_1,t, \omega-\omega_3\right) \Gamma_{\psi\zeta,ij}\left( \omega+\omega_2, \omega+\omega_4,t\right)\right.\\					
		& + 
		2 \Pi_{\psi\zeta,ji}\left(\omega,\omega-t\right) \left. \Gamma^{\times}_{\psi\zeta,ij}\left(-\omega+\omega_1, \omega-\omega_3,t\right) \Gamma^{\times}_{\psi\zeta,ji}\left( \omega+\omega_2, \omega+\omega_4,t\right) \right\} ,
	\end{split}
\end{equation}

\begin{equation}\label{eq_Xtxcmij}
	\begin{split}
		\tilde{X}^{\times}_{\psi\zeta,ij}&\left(s,t,u\right)
		=  \\			
		&\sum_{\omega} \left\{ 2  \Pi_{\psi\psi,ji}\left(\omega,\omega+s\right) \Gamma^{\times}_{\psi\zeta,ij}\left(s, \omega-\omega_{4}, \omega-\omega_{3} \right) \Gamma^{\times}_{\psi\psi,ij}\left(\omega+\omega_{1}, -\omega-\omega_{2}, s \right) \right.\\
		&-
		2 \Pi_{\psi\psi,ij}\left(\omega,\omega+s\right) \Gamma^{\times}_{\psi\zeta,ji}\left(s, \omega-\omega_{3}, \omega-\omega_{4} \right) \Gamma_{\psi\psi,ij}\left(\omega+\omega_{1}, -\omega-\omega_{2}, s \right) \\									
		&+
		2 \Pi_{\zeta\zeta,ij}\left(\omega,\omega+s\right) \Gamma_{\zeta\zeta,ij}\left( -\omega+\omega_{3},s, -\omega+\omega_{4} \right) \Gamma^{\times}_{\psi\zeta,ij}\left(s, \omega+\omega_{1}, -\omega-\omega_{2} \right) \\				
		& - 
		2 \Pi_{\psi\zeta,ji}\left(\omega,\omega-u\right) \Gamma_{\psi\zeta,ji}\left( \omega+\omega_2, \omega+\omega_3,u\right) \Gamma^{\times}_{\psi\zeta,ij}\left(-\omega+\omega_1, \omega-\omega_4, u\right)\\										
		& + 
		2 \Pi_{\psi\zeta,ij}\left(\omega,\omega-u\right) \left. \Gamma^{\times}_{\psi\zeta,ij}\left( \omega+\omega_2, \omega+\omega_3,u\right) \Gamma_{\psi\zeta,ij}\left(-\omega+\omega_1,u, \omega-\omega_4\right) \right\} .
	\end{split}
\end{equation}
\end{subequations} 
The corresponding local contributions are
\begin{subequations} 
\begin{equation}\label{eq_Xtccii}
	\begin{split}
		\tilde{X}_{\psi\psi,ii}&\left(s,t,u\right)
		=  \tilde{X}^{\times}_{\psi\psi,ii}\left(s,t,u\right) =\\			
		& + \sum_{\omega}\sum_{k} \left\{ - 2 \Pi_{\psi\psi,kk}\left(\omega,\omega-t \right) \right. \Gamma^{\times}_{\psi\psi,ik}\left( \omega-\omega_3,t,-\omega + \omega_1\right) \Gamma^{\times}_{\psi\psi,ik}\left(\omega+\omega_2,-t, -\omega-\omega_4\right) \\
		& -
		\left. \Pi_{\zeta\zeta,kk}\left(\omega,\omega-t \right) \Gamma_{\psi\zeta,ik}\left( t,\omega-\omega_3,-\omega + \omega_1\right) \Gamma_{\psi\zeta,ik}\left(-t,\omega+\omega_2, -\omega-\omega_4\right) \right\},\\		
	\end{split}
\end{equation}
\begin{equation}\label{eq_Xtmmii}
	\begin{split}
		\tilde{X}_{\zeta\zeta,ii}&\left(s,t,u\right)
		= \\			
		& \sum_{\omega}\sum_{k}\left\{ 
		- 2 \Pi_{\psi\psi,kk}\left(\omega,\omega-t\right) \Gamma_{\psi\zeta,ki}\left(-t, \omega-\omega_1, \omega-\omega_3\right) \Gamma_{\psi\zeta,ki}\left(t, \omega+\omega_4, \omega+\omega_2\right)\right.\\											
		& - 
		\Pi_{\zeta\zeta,kk}\left(\omega,\omega-t\right) \left. \Gamma_{\zeta\zeta,ki}\left( -t, -\omega+\omega_3,-\omega+\omega_1\right) \Gamma_{\zeta\zeta,ki}\left(t,-\omega-\omega_2, -\omega-\omega_4\right) \right\},\\
	\end{split}
\end{equation}
\begin{equation}\label{eq_Xtcmii}
	\begin{split}
		\tilde{X}_{\psi\zeta,ii}&\left(s,t,u\right)
		=  \tilde{X}^{\times}_{\psi\zeta,ii}\left(s,t,u\right) = \\			
		& 2 \sum_{\omega}\sum_{k} \Pi_{\psi\zeta,kk}\left(\omega,\omega-t\right) \Gamma^{\times}_{\psi\zeta,ik}\left(-\omega+\omega_1,t, \omega-\omega_3\right) \Gamma^{\times}_{\psi\zeta,ki}\left( \omega+\omega_2,t, \omega+\omega_4\right).\\
	\end{split}
\end{equation}
\end{subequations} 
\end{widetext}

\subsection{Initial conditions for the flow}\label{intialconditions}
The initial conditions for the flow of the vertices at the initial value $\Lambda = \infty$ are given by the bare vertices\cite{Kopitz2010}. However, for numerical implementations of the flow equations, the integration has to start at finite $\Lambda_i<\infty$ to be chosen much larger than all internal energy scales. For models that include a Hartree-like contribution care is needed since the contribution from the self-energy flow from $\Lambda = \infty$ to $\Lambda_i$ is non-trivial\cite{Karrasch2006,Karrasch2008}. 
To find the initial condition for the self-energy flow, we use the Schwinger-Dyson equation (see Ref.~\onlinecite{Kopitz2010}) which at large $\Lambda$ is approximated as
\begin{equation}\label{eq_schwinger_Dyson_self_energy}
\begin{split}
	\Sigma^{\Lambda}_{\psi,i}(\omega) &= \frac{1}{\beta}\sum_{\omega'}\sum_{j} 2 J_{ij}^z G^{\Lambda}_{\psi,j}(\omega').
\end{split}
\end{equation}
In analogy to Sec.~\ref{Magnetization}, the sum over the $1/\omega^\prime$ tails together with the convergence factor yield
\begin{equation}
	\Sigma^{\Lambda_i}_{\psi,i}(\omega) = \sum_{j} J_{ij}^z .
\end{equation}
This cancels the bare self-energy contribution at $\Lambda=\infty$ as read off from Eq.~\eqref{eq_action_G0}. In summary, at $\Lambda_i$, the only non-zero initial conditions are given by\footnote{Instead of carrying through the magnetic field in all calculations in Sec.~\ref{Greens_function_and_action} we could also use $h$ as initial condition for $\Gamma_{\psi,i}$ at $\Lambda \gg J,T$.}
\begin{equation}
    \Gamma_{\psi\psi,ij}^{\times}  =  -J_{ij}^z, \;\;\;  \Gamma_{\psi\zeta,ij}^{\times} =  J_{ij}^{\perp}.
\end{equation}

%%%%%%%%%%%%%%%%%%%%%%%%%%%%%%%%%%%%%%%%%%%%%%%%%%%%
\subsection{Katanin truncation} \label{Katanin}

\begin{figure}[H]
	\centering
	\includegraphics[scale=1]{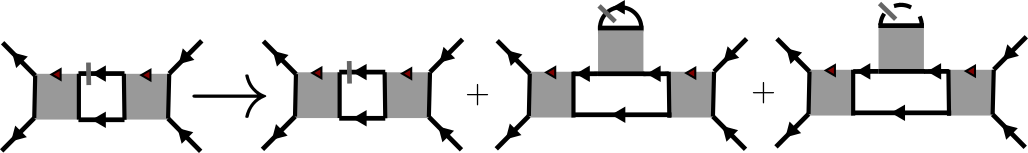}
	\caption{Katanin truncation for the vertex-flow equation by the replacement $\dot{G} \rightarrow \dot{G} + \del_{\Lambda} \Sigma$.}
	\label{fig_Katanin}
\end{figure}

So far we have neglected the six-point vertex entirely in our flow equations. However, one can include parts of it at negligible cost\cite{Katanin2004}. The prescription is to replace the single-scale propagator $\dot{G}$ in the flow equations with the sum of single-scale propagator and self-energy derivative  [cf. Fig.~\ref{fig_dia_2_mm_cc}],  $\dot{G} \rightarrow \dot{G} + \del_{\Lambda} \Sigma$. As shown in Fig.~\ref{fig_Katanin} this amounts to the net effect of the six-point vertex at least for non-overlapping loops and in lowest order perturbation theory.

%%%%%%%%%%%%%%%%%%%%%%%%%%%%%%%%%%%%%%%%%%%%%
\section{Details on numerical implementation}\label{Numerical_Implementation}

The flow equations are solved numerically with the DP5() algorithm of the julia package \emph{DifferentialEquations.jl}\cite{rackauckas2017} ODE solver with a relative and absolute accuracy goal of $10^{-6}$. Typically, we set $\Lambda_i=e^{10}$ and report the physical results at a final $\Lambda=e^{-10}$. Further, we truncate the Matsubara frequencies of the vertices at a finite Matsubara index $\pm n_{max}$. A good compromise between computation times and convergence of the results in $n_{max}$ is achieved with $n_{max} = 10$ for the bosonic Matsubara frequencies of the four-point vertices and a three times bigger value for the fermionic Matsubara frequencies of the self-energy. If the right-hand side of the flow equation requires vertices at frequencies outside of the frequency box, a projection to the last value inside the box is employed.

%%%%%%%%%%%%%%%%%%%%%%%%%%%%%%%%%%%%%%%%%%%%%
\section{Frequency symmetries of the four-point vertices}\label{Freq_syms}

To simplify the numerical treatment of the complex-valued four-point vertices $\Gamma(s,t,u)$ we use anti-symmetry under exchange of fermionic fields and the anti-unitary symmetry \eqref{eq_sym_au} which provide relations between different frequency arguments. This drastically reduces the computational cost since the left-hand side of the flow equations only needs to be evaluated for a fraction of all initially considered frequency combinations. The anti-unitary symmetry yields $\Gamma(s,t,u)=\Gamma^\star(-s,-t,-u)$ without changing the flavor or site information. This allows to restrict all vertices to, say, $s\geq 0$. In Eq.~\eqref{eq_frequencies} we display the frequency relations from (anti-)symmetry under exchange of the indices of $\Gamma_{1234}$ shown in the left column. Note that sometimes the order of the lattice sites $ij$ are swapped, this is indicated by $(i \leftrightarrow j)$.
\begin{widetext}
\begin{equation}\label{eq_frequencies}
	\begin{array}{|c||c|c|c|c|c|c|}
		\hline
		&\Gamma_{\psi\psi,ij} & \Gamma^{\times}_{\psi\psi,ij} & \Gamma_{\zeta\zeta,ij} & \Gamma_{\psi\zeta,ij} & \Gamma^{\times}_{\psi\zeta,ij} \\
		\hline
        \hline
		\Gamma_{1234} & (s,t,u) &  (s,t,u) & (s,t,u) &  
		(s,t,u) & 
		(s,t,u) \\
		\hline
		- \Gamma_{2134} &(s,-u,-t)  & & (s,-u,-t)& & \\
		\hline
		 - \Gamma_{1243} & (s,u,t) & & (s,u,t) & (s,u,t) & \\
		 \hline
		 \Gamma_{2143} & (s,-t,-u) &  \begin{array}{c}(s,-t,-u)\\\left(i\leftrightarrow j\right)\end{array} & (s,-t,-u)& & \\
		 \hline
		 \Gamma_{3412} & & & \begin{array}{c}(-s,t,-u)\\\left(i\leftrightarrow j\right)\end{array} & & \\
		 \hline
		  \Gamma_{4321} & & & \begin{array}{c}(-s,-t,u)\\\left(i\leftrightarrow j\right)\end{array} & & \\
		 	\hline
	\end{array}
\end{equation}
\end{widetext}

\bibliography{refs}

\end{document}